\documentclass[5p,twocolumn,10pt,sort,compress]{elsarticle}

\usepackage{amssymb}
 \usepackage{amssymb,amsmath,epsfig,url,multicol,multirow}
\usepackage{natbib}
\usepackage{hyperref}
\usepackage{graphicx}
\usepackage{float}
\usepackage{color}
\usepackage{diagbox}

\journal{Speech Communication}

\begin{document}

\begin{frontmatter}

\title{NHSS: A Speech and Singing Parallel Database}

\author{Bidisha Sharma\corref{cor1}}
\ead{s.bidisha@nus.edu.sg}
\author{Xiaoxue Gao}
\ead{xiaoxue.gao@u.nus.edu}
\author{Karthika Vijayan}
\ead{vijayan.karthika@nus.edu.sg}
\author{Xiaohai Tian}
\ead{eletia@nus.edu.sg}
\author{Haizhou Li}
\ead{haizhou.li@nus.edu.sg}
\address{Department of Electrical and Computer Engineering, National University of Singapore, Singapore}

\cortext[cor1]{Corresponding author}

\begin{abstract}
We present a database of parallel recordings of speech and singing, collected and released by the Human Language Technology (HLT) laboratory at the National University of Singapore (NUS), that is called NUS-HLT Speak-Sing (NHSS) database. We release this database~\footnote{\url{https://hltnus.github.io/NHSSDatabase/}} to the public to support research activities, that include, but not limited to comparative studies of acoustic attributes of speech and singing signals, cooperative synthesis of speech and singing voices, and speech-to-singing conversion. This database consists of recordings of sung vocals of English pop songs, the spoken counterpart of lyrics of the songs read by the singers in their natural reading manner, and manually prepared utterance-level and word-level annotations. The audio recordings in the NHSS database correspond to 100 songs sung and spoken by 10 singers, resulting in a total of 7 hours of audio data. There are 5 male and 5 female singers, singing and reading the lyrics of 10 songs each. In this paper, we discuss the design methodology of the database,  analyse the similarities and dissimilarities in characteristics of speech and singing voices, and provide some strategies to address relationships between these characteristics for converting one to another. We develop benchmark systems, which can be used as reference for speech-to-singing alignment, spectral mapping, and conversion using the NHSS database.
\end{abstract}


\end{frontmatter}

\vspace{-0.2cm}
\section{Introduction}
With the advent of deep learning, speech and singing technologies are often developed using neural networks that learn from  large datasets. Apart from developing technology innovation, a proper database is the basis for analyzing the characteristics of signals. Analysis of audio signals generally assists in studying the characteristics of the vocal-tract system and the excitation source signal responsible for producing them.
A fairly designed audio database can help us to solve many real-world signal processing problems and develop related realistic applications.

The widespread scope of developing technologies related to speech and singing has lead to several databases in the literature, where each of them is targeted towards a specific application. We present a brief review of some of the widespread speech and singing databases, along with the existing speech and singing parallel corpora to motivate our database development effort. 
\subsection{Speech Databases}
\footnotetext[1]{https://hltnus.github.io/NHSSDatabase/}
We would like to start with a brief review of common speech databases and the design concepts to motivate our effort to develop a speech and singing parallel database.

One broadly used database for speech signal analysis is TIMIT corpus of read speech. It is designed to provide speech data for acoustic-phonetic studies, and for the development and evaluation of automatic speech recognition (ASR) systems~\cite{garofolo1993timit}. 
In the same context, the Switchboard database is a long standing database used for speech research~\cite{godfrey1992switchboard}. Although Switchboard was originally developed for speaker recognition, it is widely used for conversational and spontaneous speech analysis, speech alignment, and spoken language understanding systems.

Among the ASR related databases, the LibriSpeech corpus~\cite{panayotov2015librispeech} is a sufficiently large publicly available database of 1,000 hours of read English speech. This corpus is extensively used in language modeling and acoustic modeling in ASR research and development~\cite{panayotov2015librispeech}. Another significant speech corpora widely used for advancing ASR research is Wall Street Journal (WSJ) CSR corpus~\cite{paul1992design}. The AMI Corpus, Aurora and TIDIGITS are other such databases available for ASR research~\cite{carletta2005ami,hirsch2000aurora,leonard1984database}.
  
Research on text-independent speaker identification is strongly supported by the international benchmarking events organized by the NIST and a large amount of data is made available in this context~\cite{martin2009nist}. VoxCeleb is a large scale audio-visual dataset~\cite{Nagrani19} used in speaker identification. The RedDots dataset is another such publicly available database~\cite{lee2015reddots}. In the context of text-dependent speaker identification systems, RSR2015 database is designed for evaluation under different duration and lexical constraints~\cite{larcher2014text}.

 The CMU ARCTIC corpus~\cite{kominek2004cmu,sharma2016speech} is one of the most popular databases for speech synthesis. The VCTK corpus~\cite{VCTK} is another popular database for multi-speaker TTS. The LJspeech~\cite{ljspeech17} speech includes 13,100 sentences, which is around 24 hours of audio and text from audiobooks read by a single speaker. The Blizzard Challenge 2011 (BC2011)~\cite{King2011TheBC}  and Blizzard Challenge 2013 (BC2013)~\cite{King2013TheBC} corpora also provide a relatively large amount of read speech from a single speaker.
 

Apart from the above mentioned databases, there are numerous other speech corpora available for the community. Broadly, the design objective of these speech databases is focused on aspects such as applicability, appropriateness, amount of speech data, time aligned transcription, speaker, and lexical variability. The speech databases can be considered as a common platform that supports decades of research in speech recognition, speaker recognition, and speech synthesis, starting from traditional hidden Markov models to deep neural networks~\cite{garofolo1993timit,martin2009nist}.

\subsection{Singing Databases}
The long standing effort in the literature to characterize singing voice and development of technologies in the field of music information retrieval (MIR) result in several singing voice databases. We review some of the largely used
databases related to different singing applications.

One widely used database in MIR is the large Dataset of synchronized Audio, LyrIcs and notes (DALI)~\cite{Meseguer-Brocal_2018}, which stands as a reference database for various music related research directions including lyrics-to-audio alignment and automatic singing voice recognition. It contains 5,358 songs with background accompaniment, time-aligned lyrics and time-aligned notes. With the increase in popularity of mobile phone karaoke applications, singing data collected from Smule'sSing! is made available for research, referred as digital  archive  of  mobile  performances (DAMP) dataset~\cite{DAMPData}. It contains more than 34K a capella (solo) singing recordings of 301 songs with textual lyrics. Gupta et al~\cite{gupta2018semi} made an effort to provide automatically time aligned lyrics for the DAMP dataset, which increases the usefulness of this large singing voice corpora.  

Another important research in MIR is singing voice extraction from background accompaniment. The iKala dataset~\cite{chan2015vocal} is specifically designed for singing voice extraction, query by humming, and melodic extraction. The music accompaniment and the singing voice are provided along with the human-labeled pitch contours and timestamped lyrics. In a similar direction, the MUSDB18 dataset~\cite{musdb18} comprises 150 full length music tracks (~10 hours duration) of different genres along with their isolated drums, bass, vocals, and other. 

The Million Song Dataset~\cite{bertin2011million} is a freely available collection of audio features and metadata for a million contemporary popular music tracks, which can be used for different applications related to the MIR research. Other singing voice databases are targeted towards specific applications like music genre classification~\cite{bogdanov2019acousticbrainz,schedl2006investigating,mckay2006large}, singer recognition~\cite{ellis2007classifying,knees2004artist,bertin2011million,Sharma2019} and  music emotion recognition~\cite{aljanaki2016studying,soleymani20131000,aljanaki2017developing}.

Apart from the above mentioned, there are several other databases with singing voice recordings. Most of these databases available for the MIR research can be used for the analysis of acoustic characteristics of singing voice, and to develop subsequent statistical models related to different technologies. 

\subsection{Speech and Singing Parallel Databases}
\begin{table} [ht]
\caption{\label{Comparison} {Comparison of speech and singing parallel databases in the literature and the proposed database (NHSS).}}
\renewcommand{\arraystretch}{1.2}
\vspace{0.2cm}
\centerline{
\scalebox{0.7}{
\begin{tabular}{|c c c c c c|}
\hline
\multicolumn {1}{|c|}{\multirow{2}{*}{\bf{Database}}} & \multicolumn{1}{|c|}{\multirow{2}{*}{\bf{NUS-48E}}} & \multicolumn{1}{|c|}{\multirow{2}{*}{\bf{SLS}}} & \multicolumn{1}{|c|}{\multirow{2}{*}{\bf{MIR-1K}}} & \multicolumn{1}{|c|}{\multirow{2}{*}{\bf{RSS}}} &\multicolumn{1}{|c|}{\bf NHSS}\\
\multicolumn {1}{|c|}{\bf{}} & \multicolumn{1}{|c|}{} & \multicolumn{1}{|c|}{} & \multicolumn{1}{|c|}{} & \multicolumn{1}{|c|}{} &\multicolumn{1}{|c|}{(\bf proposed)}\\
\hline
\hline
\multicolumn {1}{|c|}{Language} & \multicolumn{1}{|c|}{English} & \multicolumn{1}{|c|}{English} & \multicolumn{1}{|c|}{Chinese} & \multicolumn{1}{|c|}{Chinese} & \multicolumn{1}{|c|}{English}\\
\hline
\multicolumn {1}{|c|}{\#Songs} & \multicolumn{1}{|c|}{48} & \multicolumn{1}{|c|}{90} & \multicolumn{1}{|c|}{110} & \multicolumn{1}{|c|}{10} &  \multicolumn{1}{|c|}{100}\\
\hline
\multicolumn {1}{|c|}{\#Singers} & \multicolumn{1}{|c|}{12} & \multicolumn{1}{|c|}{10} & \multicolumn{1}{|c|}{19} & \multicolumn{1}{|c|}{20} & \multicolumn{1}{|c|}{10}\\
\hline
\multicolumn {1}{|c|}{Duration (min)} & \multicolumn{1}{|c|}{169} & \multicolumn{1}{|c|}{560} & \multicolumn{1}{|c|}{133} & \multicolumn{1}{|c|}{172} & \multicolumn{1}{|c|}{421}\\
\hline
\multicolumn {1}{|c|}{Manual alignment} & \multicolumn{1}{|c|}{Yes} & \multicolumn{1}{|c|}{No} & \multicolumn{1}{|c|}{No} & \multicolumn{1}{|c|}{No} & \multicolumn{1}{|c|}{Yes}\\
\hline
\multicolumn {1}{|c|}{Professional singers} & \multicolumn{1}{|c|}{No} & \multicolumn{1}{|c|}{Yes} & \multicolumn{1}{|c|}{No} & \multicolumn{1}{|c|}{No} & \multicolumn{1}{|c|}{Yes}\\
\hline
\end{tabular}}
}
\end{table}

For applications like speech-to-singing conversion, speaker identity conversion, and singing voice conversion, a comparative study of speech and singing voice is required. This necessitates a multi-speaker speech and singing voice parallel database.

Despite many attempts, there are very few databases to support an effective comparative study of speech and singing voice. The NUS sung and spoken lyrics (NUS-48E) corpus~\cite{duan2013nus} provides speech and singing parallel recordings. It is a phonetically annotated corpus with 48 English songs. The lyrics of each song are sung and spoken by 12 subjects representing a variety of voice types and accents. There are 20 unique songs, each of which is covered by at least one male and one female subject, with a total duration of 169 mins. Although the NUS-48E corpus is annotated in phone level, due to the very short length of phones and the presence of co-articulation effect specifically in singing, the manually identified phone boundaries may not be reliable. This corpus is also of limited size which restricts its applicability in developing statistical models or mapping schemes for related applications.  

The MIR-1K database~\cite{hsu2009improvement} consists of 1000 song clips, which are extracted from 110 recordings of Chinese pop songs sung by amateurs singers. This database also provides human-labeled pitch values, unvoiced sounds and vocal/nonvocal segments, lyrics, and the speech recordings of the lyrics for each clip. The MIR-1K database contains total 133 mins of sung audio along with music accompaniment in separate channels and is specifically designed for the research of singing voice separation. As the singing vocals in the MIR-1K~\cite{hsu2009improvement} database are not sung by professional singers, this database may not be suitable for the speech-to-singing conversion research.

Another speech and singing parallel database in Chinese language, referred to as RSS database has been published in~\cite{shi2019addressing, shi_2019_3241566}. The RSS database consists of parallel speech and singing data for 10 male and 10 female singers and is used in text-dependent speaker verification. However, this database does not provide any transcription or word/phoneme boundary information.

A primitive version of the database introduced in this paper is presented in~\cite{gao2018nus}, and referred as NUS-HLT Spoken Lyrics and Singing (SLS) corpus. The SLS corpus is the collection of the recordings of speech and singing for 100 songs without any manual intervention and annotation. A comparison of different speech and singing parallel databases is shown in Table~\ref{Comparison}.

A relatively large, well designed, multi-speaker, uniform, and precisely time-aligned publicly available speech and singing parallel database has crucial importance in the research of MIR. It would allow for a comparative, quantitative, and systematic study between speech and singing, and application-oriented statistical modeling.

\subsection{Motivation}

Speaking and singing voices are produced from the human vocal-tract system excited by the glottal excitation signal. Although both speaking and singing voices are generated by the same production system, they tend to produce two very different kinds of  signals \cite{II-3}. The major differences between speaking and singing voices lie in the loudness, pitch, duration and formant variations \cite{II-3, fujisaki1983dynamic, KTHSundberg}. Studying parallel utterances of spoken and sung voices can shed light on the different configurations of the vocal-tract system and glottal excitation that rendered these two kinds of signals, even though the linguistic content remains the same. A coherent understanding of the human voice production mechanism behind speaking and singing can benefit numerous applications such as singing voice synthesis, conversion of voice identity of one singer to another, conversion of read lyrics to singing. These applications are under rapid development and soaring demand from the entertainment industry, which necessitate a well designed speech and singing parallel database~\cite{vijayan2018speech}.


In order to demonstrate the prevalent significance of a sufficiently large, manually annotated speech and singing parallel database on technology development, we broadly discuss the  speech-to-singing conversion and its sub-components. Speech-to-singing conversion is the process of converting natural speech to singing with the same lyrical content, by retaining the speaker characteristics and correct prosody structure as that of a high quality song. This technology allows the user to synthesize personalized singing just by reading lyrics of a song, which is very appealing to the general public~\cite{dong2014i2r,cen2012template,Gupta2019NUSS}.

The basic aspect of speech-to-singing conversion is to modify the prosodic features of user's speech, with respect to reference singing, which is then used to generate the singing voice by retaining the user's voice identity. To convert user's speech to appropriate singing, the mapping of prosodic characteristics from speech to singing is very crucial. This mapping can be performed with alignment information between frames of speech and singing, which we refer to as speech-to-singing alignment. The quality of synthesized singing depends on the accuracy of the alignment to a significant extent. The alignment strategies require a comparative study between speech and singing. Despite several differences between speech and singing, such as duration, pitch, spectral energy content, the correct alignment can be achieved by comparing commonalities between the two. To study the speaker independent commonalities between speech and singing and evaluate the efficacy of such a temporal alignment method, a multi-speaker speech and singing parallel database is of crucial importance. 

Due to the prominent difference in the formant characteristics of speech and singing spectrum, the spectral transformation/mapping techniques are proposed to achieve better naturalness in the converted singing voice~\cite{gao2019speaker}, while retaining the speaker information. This spectral mapping also necessitates a speech and singing parallel database with a sufficient amount of audio data. 

Furthermore, the unique features in singing vocals~\cite{sundberg1990science,dayme2009dynamics} and their perceptual effects~\cite{saitou2004analysis,ohishi2006human} are important research directions in the field of speech-to-singing conversion. The singing voice recorded from professional singers can also be used as the reference singing template for prosody generation, to apply in singing voice synthesis. 

To precisely evaluate the effectiveness of an existing speech-to-singing conversion method, we can compare the converted singing obtained from the user's speech to the user's original singing. Therefore, evaluation of singing voice synthesis or singing voice conversion can be largely benefited from such a database.

Development of the above mentioned efficient statistical and adaptive systems requires databases of speech and singing parallel utterances. A database with professional singers' singing and corresponding speech for multiple songs would facilitate the research in speech-to-singing conversion in terms of reference prosody generation, speech and singing alignment, spectral mapping, conversion methodology, and evaluation. As mentioned earlier, there are very few publicly available databases in this area to advance the research based on a common platform. 
 
With this outlook of the importance of a speech and singing parallel database, we propose the NHSS database, which consists of spoken and sung lyrics of English popular songs along with their manually labeled utterance- and word-level annotations. In this work, we also conduct studies on speech-to-singing alignment, spectral mapping, and speech-to-singing conversion, where we aim to use existing methods to develop reference systems and report the findings. With the demonstration of these applications, we validate the usability of the NHSS database. This work may serve as a common platform for various applications that require parallel speech and singing recordings with the same lyrical content.

In~\cite{gao2018nus}, the ongoing effort to collect and design the NHSS database is explained briefly from the perspective of its significance in different research areas. The work presented here is significantly different from~\cite{gao2018nus}, in terms of the design of the complete database with manual intervention, time aligned word/utterance annotation, detailed statistical analysis, and evaluation of the related baseline systems, which can be seen from Table~\ref{Comparison}. 

The rest of the paper is organized as follows. In Section~\ref{Sec2}, we explain the collection, processing, and organization of the NHSS database. The acoustic analysis and comparison of characteristics of speech and singing are described in Section~\ref{Sec3}. The methods for speech-to-singing alignment with respect to the NHSS database are presented and evaluated in Section~\ref{Sec4}. In Section~\ref{Sec5}, we present different approaches for spectral mapping followed by the development of speech-to-singing conversion systems. Finally, we draw conclusions of this work in Section~\ref{Sec8}.

\section{NUS-HLT Speak-Sing (NHSS) Database}\label{Sec2}
The NHSS database consists of 100 songs from 10 professional singers, each of them singing and speaking 10 songs. The database provides 4.75 hours of sung and 2.25 hours of spoken data. In total, we have 7 hours of audio data with manually annotated utterance and word boundaries.

In the design of this database, we emphasize several factors such as audio quality, singer selection, one-to-one correspondence of words/utterances between spoken and sung audio, accurate word boundary, usability for various applications, and  organization. 
To achieve these objectives, and ensure quality, we pay special attention to data collection and data processing stages. We briefly explain each of these in the following sections. We also present the statistics and  organization of the database in the subsequent sections.

\subsection{Data Collection}
\begin{table} [ht]
\caption{\label{Songs} {English song list with year of release included in the NHSS database.}}
\renewcommand{\arraystretch}{1}
\vspace{0.2cm}
\centerline{
\scalebox{0.85}{
\begin{tabular}{|c c|}
\hline
\multicolumn {1}{|c|}{\bf{Song ID}} & \multicolumn{1}{|c|}{\bf{Song title}} \\
\hline
\hline
\multicolumn {1}{|c|}{S01} & \multicolumn{1}{|c|}{Billie Jean}\\
\hline
\multicolumn {1}{|c|}{S02} & \multicolumn{1}{|c|}{Bridge Over Troubled Water}\\
\hline
\multicolumn {1}{|c|}{S03} & \multicolumn{1}{|c|}{Can't Help Falling in Love}\\
\hline
\multicolumn {1}{|c|}{S04} & \multicolumn{1}{|c|}{Dancing Queen}\\
\hline
\multicolumn {1}{|c|}{S05} & \multicolumn{1}{|c|}{Hey Jude}\\
\hline
\multicolumn {1}{|c|}{S06} & \multicolumn{1}{|c|}{How Deep is Your Love}\\
\hline
\multicolumn {1}{|c|}{S07} & \multicolumn{1}{|c|}{How Do I Live}\\
\hline
\multicolumn {1}{|c|}{S08} & \multicolumn{1}{|c|}{I'll Be There}\\
\hline
\multicolumn {1}{|c|}{S09} & \multicolumn{1}{|c|}{Let it be}\\
\hline
\multicolumn {1}{|c|}{S10} & \multicolumn{1}{|c|}{Linger}\\
\hline
\multicolumn {1}{|c|}{S11} & \multicolumn{1}{|c|}{My Heart Will Go On}\\
\hline
\multicolumn {1}{|c|}{S12} & \multicolumn{1}{|c|}{Poker Face}\\
\hline
\multicolumn {1}{|c|}{S13} & \multicolumn{1}{|c|}{Stand by Me}\\
\hline
\multicolumn {1}{|c|}{S14} & \multicolumn{1}{|c|}{Staying Alive}\\
\hline
\multicolumn {1}{|c|}{S15} & \multicolumn{1}{|c|}{Take My Breath Away}\\
\hline
\multicolumn {1}{|c|}{S16} & \multicolumn{1}{|c|}{Total Eclipse of the Heart}\\
\hline
\multicolumn {1}{|c|}{S17} & \multicolumn{1}{|c|}{Yesterday}\\
\hline
\multicolumn {1}{|c|}{S18} & \multicolumn{1}{|c|}{You Light Up My Life}\\
\hline
\multicolumn {1}{|c|}{S19} & \multicolumn{1}{|c|}{Foolish Games}\\
\hline
\multicolumn {1}{|c|}{S20} & \multicolumn{1}{|c|}{I Will Always Love You}\\
\hline
\end{tabular}}
}
\end{table}


\begin{table} [ht]
\caption{\label{SongsSinger} {Collection of song IDs sung/spoken by each singer in the NHSS database.}}
\renewcommand{\arraystretch}{1.2}
\vspace{0.2cm}
\centerline{
\scalebox{0.85}{
\begin{tabular}{|c c|}
\hline
\multicolumn {1}{|c|}{\bf{Singer ID}} & \multicolumn{1}{|c|}{\bf{Song IDs}}\\
\hline
\hline
\multicolumn {1}{|c|}{F01} & \multicolumn{1}{|c|}{S01, S03, S04, S05, S06, S06,  S09, S11, S12, S15}\\
\hline
\multicolumn {1}{|c|}{F02} & \multicolumn{1}{|c|}{S04, S05, S06, S07, S09, S11, S15, S16, S17, S18}\\
\hline
\multicolumn {1}{|c|}{F03} & \multicolumn{1}{|c|}{S03, S04, S05, S06, S08, S09, S16, S18, S19, S20}\\
\hline
\multicolumn {1}{|c|}{F04} & \multicolumn{1}{|c|}{S03, S04, S05,S06, S09, S11, S12, S15, S16, S18}\\
\hline
\multicolumn {1}{|c|}{F05} & \multicolumn{1}{|c|}{S03, S05, S06, S08, S09, S10, S11, S12, S15, S18}\\
\hline
\multicolumn {1}{|c|}{M01} & \multicolumn{1}{|c|}{S01, S03, S04, S05, S06, S09,  S13, S14, S16, S17}\\
\hline
\multicolumn {1}{|c|}{M02} & \multicolumn{1}{|c|}{S04, S05, S06, S06, S09, S10,  S12, S15, S17, S18}\\
\hline
\multicolumn {1}{|c|}{M03} & \multicolumn{1}{|c|}{S03, S05, S06, S09, S10, S11, S13, S15, S17, S18}\\
\hline
\multicolumn {1}{|c|}{M04} & \multicolumn{1}{|c|}{S03, S05, S06, S09, S10, S11, S12, S14, S15, S17}\\
\hline
\multicolumn {1}{|c|}{M05} & \multicolumn{1}{|c|}{S02, S03, S05, S07, S09, S10, S11, S12, S15, S18}\\
\hline
\end{tabular}}
}
\end{table} 
We recruit professional singers who either have a diploma in vocal training or have experience of more than three years in public singing. The singers are selected through an audition process. We ensure that the singers can speak and sing in English without prominent mother tongue influence. We select 10 singers (5 male and 5 female), each of them singing/speaking 10 songs according to their singing convenience (total 100 songs), from a song list consisting of 20 unique English songs shown in Table~\ref{Songs}. We select classic songs, so that they are well known across a wide range of population.

Before the actual recording of the songs, we have provided the original songs to the singers for practicing and listening. During the recording of singing vocals, the background accompaniment of the original song is played through headphones to the singers, and they are asked to sing in synchronization and the same key with the background accompaniment. They also need to maintain the same tempo, pitch, and lyrics to mimic the original singer. The singing vocals are verified by experts and re-recorded if there is any deviation from the original singing.  The recording of the speech is performed by instructing the singers to read the lyrics of songs in their natural speaking manner, without taking long pauses in between. All recordings are performed with 48kHz sampling rate and 16 bits per sample.

In Table~\ref{SongsSinger}, we list down the 10 singers in our database and the songs each of them recorded. We denote the female singers as F01, F02,...F05, and male singers as M01, M02,...M05.

We record the sung and spoken audio of these songs in a  studio environment using {\it Neumann U87 microphone} and {\it RME Fireface UC SoundCard}, under the supervision of a trained and experienced sound engineer. 

\subsection{Data Processing} \label{DataProcessing}
   
We process the collected audio data to further refine and obtain the utterance- and word-level annotations for both speech and singing. In order to ensure the preciseness of the word boundaries, we employ both automatic annotation and manual processing, followed by a post-processing stage to deal with the erroneous recordings. 

\begin{figure}[t!]
\vspace{-0.5cm}
\centering
\centerline{\epsfig{figure=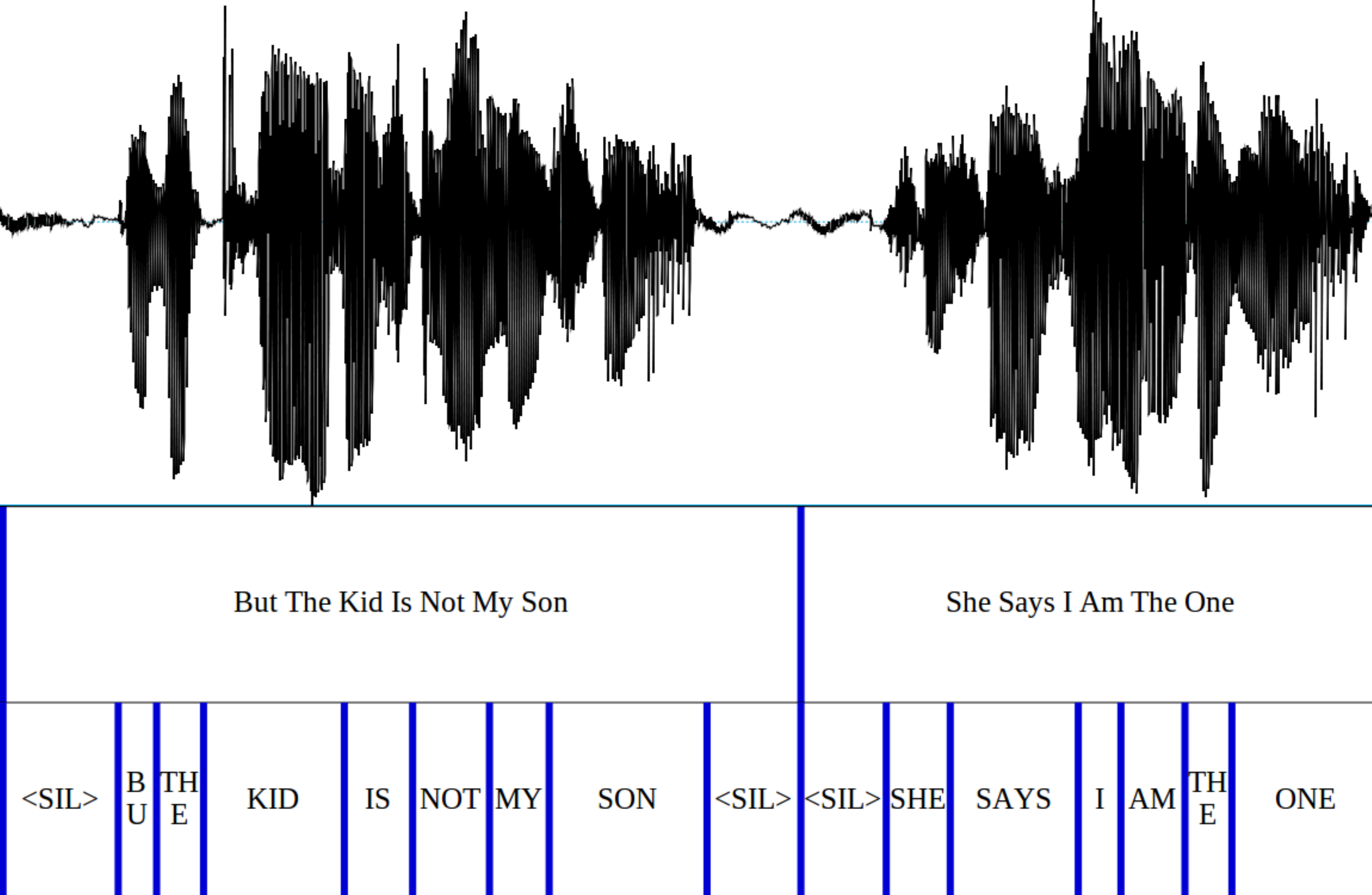,height=1.2in, width=2.4in}}
\vspace{-0.2cm}
\caption{Illustration of manual processing in Praat.}
\label{SpeechPraat}
\end{figure}

\subsubsection{Automatic Annotation}
We perform automatic annotation for both speech and singing audio with respect to the corresponding lyrics to obtain initial word boundaries for all 100 songs in the NHSS database.

We use Montreal forced aligner (MFA)~\cite{mcauliffe2017montreal,Montreal} to obtain word boundaries by aligning speech utterances with corresponding lyrics. MFA uses a standard Gaussian mixture model (GMM)/hidden Markov model (HMM) architecture, that is built with Kaldi recipes~\cite{povey2011kaldi,panayotov2015librispeech}. To build such a model, monophone GMMs are trained in an iterative manner and used to generate a basic alignment. Then, the triphone GMMs are trained to take surrounding phonetic context into account, along with the clustering of triphones to combat sparsity, which are used to generate the alignments. The triphone models are then used for learning acoustic feature transforms on a per-speaker basis, to make the models more applicable to speakers in other databases. In this work, we use the pre-trained English acoustic models available in~\cite{Montreal}, trained on the LibriSpeech corpus~\cite{panayotov2015librispeech}. 

In view of the differences between speech and singing, it is not appropriate to use the same speech models to align singing vocals to the lyrics. We adapt the acoustic model of speech to  singing  voice using  speaker  adaptation methods~\cite{mesaros2010automatic,gupta2018semi}. 

The baseline speech acoustic model is a tri-phone GMM-HMM trained on the  LibriSpeech corpus~\cite{panayotov2015librispeech} using mel frequency cepstral coefficients (MFCC) features on Kaldi toolkit~\cite{povey2011kaldi}. To adapt these speech models to the singing voice, a semi-supervised speaker adaptive training method is applied, with the lyrics-aligned DAMP solo-singing dataset \cite{gupta2018semi, DAMPData}. The feature-space maximum likelihood linear regression is used for the adaptation of the
speech models. These singing adapted models can be applied to lyrics-to-audio alignment~\cite{BidishaICASSP}. We use these models to obtain the initial word boundaries of the singing utterances. 

\subsubsection{Manual Processing} \label{ManualProcessing}

\begin{table} [h!]
\vspace{-0.5cm} 
\caption{\label{InterTrans} {Inter-annotator deviation for singing and speech.}}
\renewcommand{\arraystretch}{1}
\vspace{.3cm}
\centerline{
\scalebox{0.9}{
\begin{tabular}{|c c c|}
\hline
\multicolumn{3}{|c|}{\bf{Boundary deviation (msec)}}\\
\hline
\multicolumn{1}{|c|}{{}} & \multicolumn{1}{|c|}{{Word}} & \multicolumn{1}{|c|}{{Utterance}}\\
\hline
\hline
\multicolumn{1}{|c|}{{Singing}} & \multicolumn{1}{|c|}{{8.70}} & \multicolumn{1}{|c|}{{46.08}}\\
\hline
\multicolumn{1}{|c|}{{Speech}} & \multicolumn{1}{|c|}{{5.70}} & \multicolumn{1}{|c|}{{29.00}}\\
\hline
\end{tabular}}
}
\end{table} 

\begin{table} [h!]
\vspace{-0.4cm} 
\caption{\label{ManualAlignmentSpeech} {Comparison of utterance and word boundaries obtained from automatic and manual annotations for singing and speech.}}
\renewcommand{\arraystretch}{1}
\vspace{.3cm}
\centerline{
\scalebox{0.9}{
\begin{tabular}{|c c c|}
\hline
 \multicolumn{3}{|c|}{\bf{Boundary deviation (msec)}}\\
 \hline
\multicolumn{1}{|c|}{\bf{}} & \multicolumn{1}{|c|}{\bf{Word}} & \multicolumn{1}{|c|}{\bf{Utterance}}\\
\hline
\hline
\multicolumn{1}{|c|}{Singing} & \multicolumn{1}{|c|}{103.90} & \multicolumn{1}{|c|}{185.30}\\
\hline
\multicolumn{1}{|c|}{Speech} & \multicolumn{1}{|c|}{25.30} & \multicolumn{1}{|c|}{87.80}\\
\hline
\end{tabular}}
}
\end{table} 

\begin{table} [h!]
\vspace{-0.4cm} 
\caption{\label{Errors} {Statistics of different types of errors detected in manual processing.}}
\renewcommand{\arraystretch}{1.4}
\vspace{.2cm}
\centerline{
\resizebox{0.48\textwidth}{!}{
\begin{tabular}{|c c c c c c c|}
\hline
\multicolumn {1}{|c|}{\multirow{3}{*}{\bf{Error type}}} & \multicolumn {2}{|c|}{\bf{Entire database}} & \multicolumn{2}{|c|}{\bf{Male}} & \multicolumn{2}{|c|}{\bf{Female}}\\
\cline{2-7}
\multicolumn{1}{|c|}{} & \multicolumn{1}{|c|}{\multirow{2}{*}{\#Utterances}} & \multicolumn{1}{|c|}{Duration} & \multicolumn{1}{|c|}{\multirow{2}{*}{\#Utterances}} & \multicolumn{1}{|c|}{Duration} & \multicolumn{1}{|c|}{\multirow{2}{*}{\#Utterances}} & \multicolumn{1}{|c|}{Duration}\\
\multicolumn{1}{|c|}{} &  & \multicolumn{1}{|c|}{(sec)} &  & \multicolumn{1}{|c|}{(sec)} &  & \multicolumn{1}{|c|}{(sec)}\\
\hline
\hline
\multicolumn{1}{|c|}{Deletion} & \multicolumn{1}{|c|}{102} & \multicolumn{1}{|c|}{510.83} & \multicolumn{1}{|c|}{55} & \multicolumn{1}{|c|}{217.36} & \multicolumn{1}{|c|}{47} & \multicolumn{1}{|c|}{293.47}\\
\hline
\multicolumn{1}{|c|}{Insertion} & \multicolumn{1}{|c|}{21} & \multicolumn{1}{|c|}{108.74} & \multicolumn{1}{|c|}{11} & \multicolumn{1}{|c|}{40.47} & \multicolumn{1}{|c|}{10} & \multicolumn{1}{|c|}{68.27}\\
\hline
\multicolumn{1}{|c|}{Substitution} & \multicolumn{1}{|c|}{146} & \multicolumn{1}{|c|}{776.12} & \multicolumn{1}{|c|}{62} & \multicolumn{1}{|c|}{281.39} & \multicolumn{1}{|c|}{84} & \multicolumn{1}{|c|}{494.73}\\
\hline
\multicolumn{1}{|c|}{Total} & \multicolumn{1}{|c|}{269} & \multicolumn{1}{|c|}{23.26 min} & \multicolumn{1}{|c|}{128} & \multicolumn{1}{|c|}{8.99 min} & \multicolumn{1}{|c|}{141} & \multicolumn{1}{|c|}{14.27 min}\\
\hline
\end{tabular}}
}
\end{table}

\begin{table*} [ht]
\vspace{-0.2cm} 
\caption{\label{Overview} {A summary of the statistics of the NHSS database.}}
\renewcommand{\arraystretch}{1.2}
\vspace{0.2cm}
\centerline{
\scalebox{0.85}{
\begin{tabular}{|c c c c c c c|}
\hline
\multicolumn {1}{|c|}{\bf{Parameter}} & \multicolumn{2}{|c|}{\bf{All data}} & \multicolumn{2}{|c|}{\bf{Male singers' data}}  & \multicolumn{2}{|c|}{\bf{Female singers' data}}\\
\cline{2-7}
\multicolumn {1}{|c|}{} & \multicolumn{1}{|c|}{\bf{Singing}} & \multicolumn{1}{|c|}{\bf{Speech}}  & \multicolumn{1}{|c|}{\bf{Singing}} & \multicolumn{1}{|c|}{\bf{Speech}}  & \multicolumn{1}{|c|}{\bf{Singing}} & \multicolumn{1}{|c|}{\bf{Speech}}\\
\hline
\hline
\multicolumn{1}{|c|}{\#Singers/speakers}& \multicolumn{2}{|c|}{10} & \multicolumn{2}{|c|}{5} & \multicolumn{2}{|c|}{5}\\
\hline
\multicolumn{1}{|c|}{\#Songs}& \multicolumn{2}{|c|}{100} & \multicolumn{2}{|c|}{50} & \multicolumn{2}{|c|}{50}\\
\hline
\multicolumn{1}{|c|}{\#Unique songs}& \multicolumn{2}{|c|}{20} & \multicolumn{2}{|c|}{17} & \multicolumn{2}{|c|}{17}\\
\hline
\multicolumn{1}{|c|}{\#Utterances}& \multicolumn{2}{|c|}{2,791} & \multicolumn{2}{|c|}{1,436} & \multicolumn{2}{|c|}{1,355}\\
\hline
\multicolumn{1}{|c|}{\#Words}& \multicolumn{2}{|c|}{20,408} & \multicolumn{2}{|c|}{10,491} & \multicolumn{2}{|c|}{9,917}\\
\hline
\multicolumn{1}{|c|}{\#Unique words}& \multicolumn{2}{|c|}{663} & \multicolumn{2}{|c|}{579} & \multicolumn{2}{|c|}{581}\\
\hline
\multicolumn{1}{|c|}{Average utterance duration}& \multicolumn{1}{|c|}{6.13 sec} & \multicolumn{1}{|c|}{2.93 sec} & \multicolumn{1}{|c|}{6.04 sec} & \multicolumn{1}{|c|}{2.82 sec} & \multicolumn{1}{|c|}{6.23 sec} & \multicolumn{1}{|c|}{3.04 sec}\\
\hline
\multicolumn{1}{|c|}{Average word duration} & \multicolumn{1}{|c|}{729 msec} & \multicolumn{1}{|c|}{295 msec} & \multicolumn{1}{|c|}{709 msec} & \multicolumn{1}{|c|}{277 msec} & \multicolumn{1}{|c|}{751 msec} & \multicolumn{1}{|c|}{316 msec}\\
\hline
\multicolumn{1}{|c|}{Total duration}& \multicolumn{1}{|c|}{285.24 min} & \multicolumn{1}{|c|}{136.12 min} & \multicolumn{1}{|c|}{144.62 min} & \multicolumn{1}{|c|}{67.53 min} & \multicolumn{1}{|c|}{140.61 min} & \multicolumn{1}{|c|}{68.59 min}\\
\hline
\end{tabular}}
}
\vspace{-.3cm}
\end{table*} 
The automatically obtained word boundaries for both speech and singing are further subjected to manual correction. To ensure correct word boundaries and one-to-one correspondence between words/utterances in spoken and sung audio, we perform two-level manual processing. 

In the first level, the two annotators ($\text{Team}_1$) are trained to correct automatically obtained word boundaries by listening to and observing the spectrogram of each word using Praat~\cite{Praat}. For each of the 100 songs, we have spoken and sung audio with corresponding label files. An example of manual processing in Praat is shown in Figure~\ref{SpeechPraat}, with utterance- and word-level panels for annotation. Each of the two annotators performs manual processing for 50 songs corresponding to 5 singers. Each song is corrected by one subject only. 
  
Apart from the annotation of word boundaries, the annotators also compare the number of words and utterances, assuring that there is a one-to-one correspondence between speech and singing. If a word is present in the audio and not present in the text, they mark the corresponding label and vice-versa as {\it word$<$missing$>$} and {\it word$<$ins$>$}, respectively. If a word present in the audio is different than that in the text, they mark the corresponding label as {\it word$<$sub$>$}.
 
The corrected word boundaries are further verified in the second level of manual processing by two different annotators ($\text{Team}_2$). Unlike the first level manual processing, in this case, all 100 songs are corrected by both annotators.

We select the reliable annotator by manual intervention, where we first find the deviation in word boundaries annotated by the two annotators. We manually check word boundaries where the deviation between the two annotators is more, and the annotator with maximum correct boundaries is noted as a reliable annotator. We consider the boundaries corrected by the reliable annotator as final word boundaries to be included in the NHSS database. We note that the spoken and sung versions of a song are annotated by the same annotator. In Table~\ref{InterTrans}, we show the average word and utterance boundary deviation between the annotations provided by both the annotators. The inter-annotator word boundary deviation is 8.70 msec for singing and 5.70 msec for speech.

In Table~\ref{ManualAlignmentSpeech} we show the average word and utterance boundary deviation between the word boundaries obtained from automatic and manual annotations for both singing and speech. For speech, the average word boundary deviation is 25.30 msec, while for singing the same is 103.90 msec. This in turn shows better compatibility of the MFA. The utterance boundary deviation is comparatively large in both speech and singing, because of the long silence segments present at the beginning and end of the utterances.
   
\subsubsection{Post-Processing}

To ensure further correctness after manual processing, the post-processing stage is crucial. In the post-processing stage we analyse different types of errors, compare the number of words/utterances and one-to-one correspondence between speech and singing.

As mentioned in Section~\ref{ManualProcessing}, in the manual processing the annotators identify deletion, insertion, and substitution errors while matching the speak-sing pairs. In the post-processing stage, we exclude the utterance pairs with these errors and we end up with 2,791 utterance pairs annotated in the final database. In Table~\ref{Errors}, we show the number of utterances and duration affected due to such errors for the entire database. In total, we discard 269 utterances with a duration of 23.26 mins (including both speech and singing audio) from the database due to such errors. 

Another important aspect of post-processing is to ensure one-to-one correspondence of all utterances/words between speech and singing. We believe that the post-processed database is accurate in terms of utterance and word boundaries, and error free. 

\vspace{-0.2cm}
\subsection{Statistics}
\label{sec:stats}
\begin{figure}[h!]
\centering
\centerline{\epsfig{figure=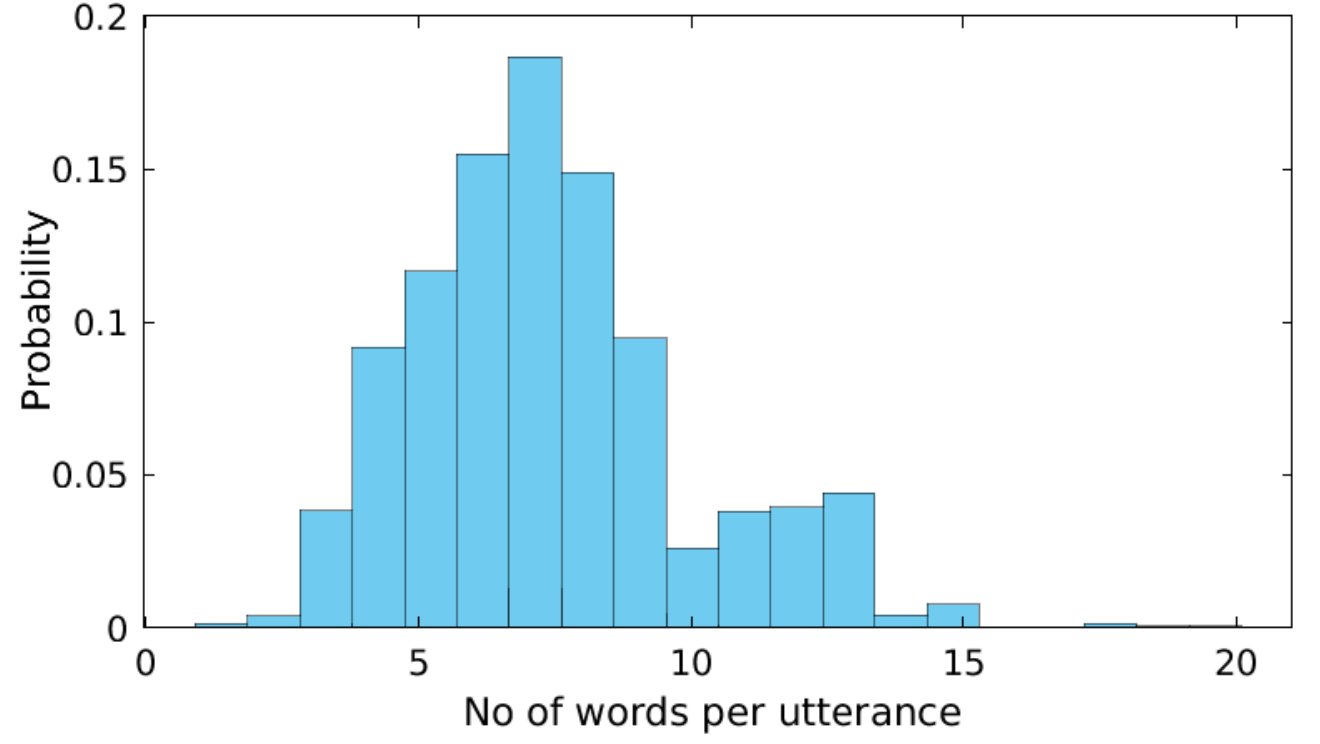,height=1.8 in, width=3in}}
\vspace{-0.4cm}
\caption{Histogram of number of words per utterance for the NHSS database.}
\label{NoOfUtt}
\vspace{-0.2cm}
\end{figure}

In this section, we give a summary of the NHSS database in terms of the number of utterances, words, and their durations. In Table~\ref{Overview}, we show the statistics of the entire database along with male and female singers' data.
As shown in Table~\ref{Overview}, the database contains audio recordings of 10 singers, a total of 100 songs, and 20 unique songs. It contains 2,791 utterances and 20,408  words, for each singing and speech. In total, the database has 5,582 utterances and 40,816 words, with 663 unique words. The duration of sung and spoken audio in the NHSS database is 285.24 mins and 136.12 mins, respectively, resulting in a total of 421.36 mins (7 hours) of audio data.


We observe uniformity in average word duration and total duration between male and female singers'/speakers' data. The total duration for male singing and speech is 144.62 mins and 67.53 min, respectively; the counterpart for female singing and speech audio is 140.61 mins and 68.59 min, respectively. In Figure~\ref{NoOfUtt}, we show the histogram distribution of the number of words per utterance. We observe that in most of the utterances we have 6 to 8 words. 


\begin{figure*}[t!]
\centering
\centerline{\epsfig{figure=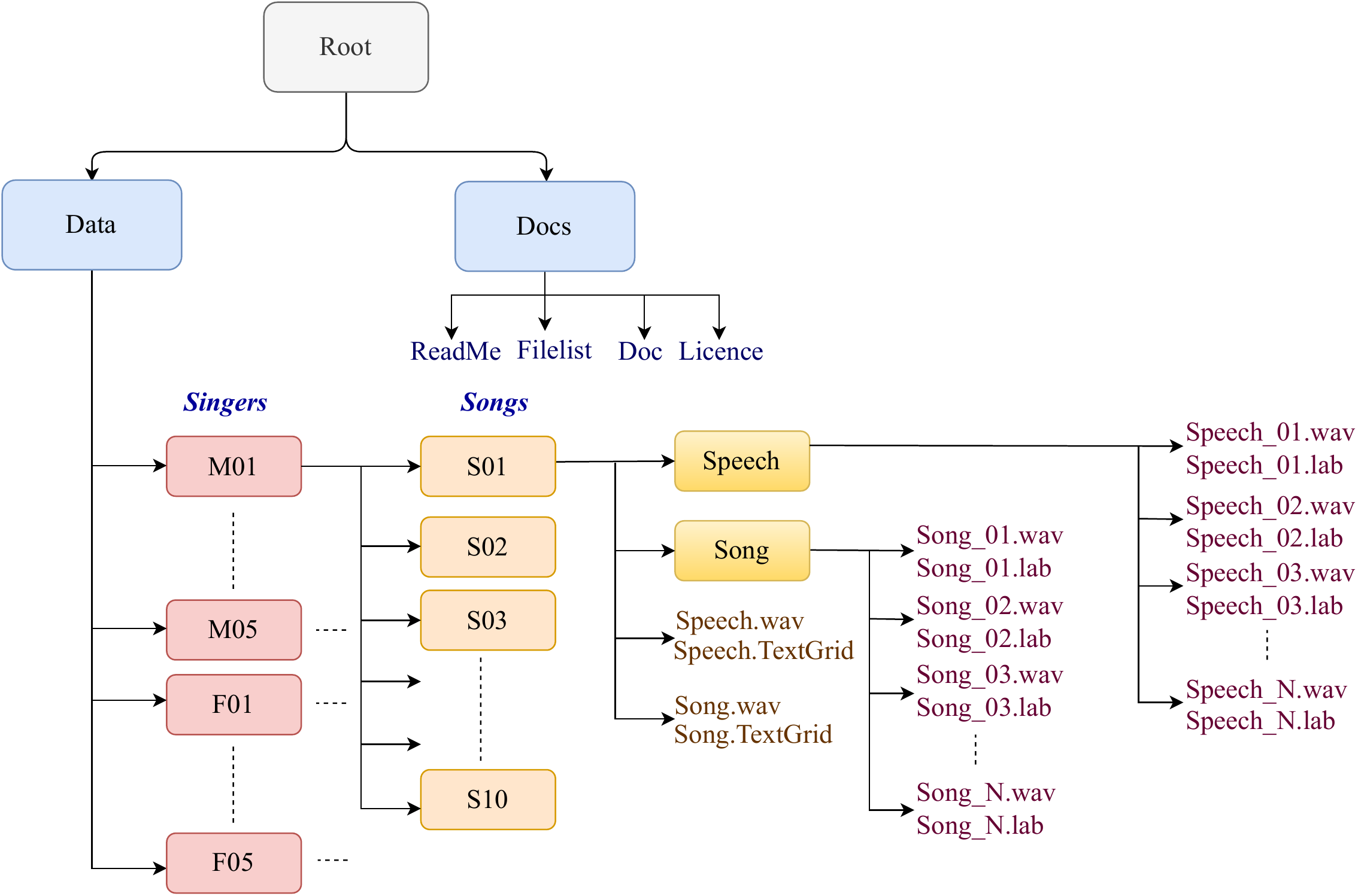,height=3 in, width=5in}}
\vspace{-0.2cm}
\caption{Schematic representation of directory organization of the NHSS database.}
\label{SpeechSingingCorpus}
\vspace{-0.2cm}
\end{figure*}

\subsection{Organization}

The organization of the NHSS database is shown in Figure~\ref{SpeechSingingCorpus}. Under the {\it Root} directory, we have two folders {\it Data} and {\it Docs}. Under {\it Docs}, we provide ReadMe, Filelist, Doc and Licence. Inside the Data folder, we have 10 sub-folders for 10 singers. The 5 male singers are represented as M01, M02,..M05 and the 5 female singers are represented as F01, F02,..F05. The 20 unique songs are represented as S01, S02,...S20.  In each singer's folder, we have 10 sub-folders corresponding to  10 songs, where the folder name represents the song ID. Each song folder comprises 6 components, which are {\it Speech.wav},  {\it Song.wav}, {\it Speech.TextGrid}, {\it Song.TextGrid}, and two sub-folders, namely {\it Speech} \& {\it Song}. {\it Speech.wav},  {\it Song.wav}, {\it Speech.TextGrid}, {\it Song.TextGrid} represent audio files and label files for speech and singing of each song, respectively. The folders {\it Speech} and {\it Song} contain the wavfiles (.wav) and labfiles (.lab) corresponding to the split utterances of spoken and sung versions of each song.

\section{Analysis of Acoustic Properties}\label{Sec3}
In this section, we discuss and analyse different acoustic characteristics of speech and singing signals with reference to the NHSS database.

Singing is a form of art that is rich in the expression of emotions, providing exciting and soothing feelings to listeners. The expressiveness in singing is generally delivered by singers using a  wide range of variations in energy, duration, pitch, and spectrum. A trained singer is able to maneuver the subglottal pressure efficiently, thereby controlling the rate of airflow at the glottis to cater to the required loudness and pitch for the melody, content, and genre of the song \cite{I-5, I-69}. The spectrum in singing voice gets modulated by the pitch, which makes it different from the speech spectrum. Furthermore, these observations in the literature can be quantified and subsequently applied in technology development, provided we have speech and singing audio signals for the same linguistic content.

Parallel utterances make a comparative study between speech and singing possible. The NHSS database features English pop songs with parallel speech and singing audio, which is useful in the investigation of the properties of the vocal-tract system and excitation signal. Some of these properties stay invariant, while some others may vary while producing the same sound in speech and singing. Furthermore, the clean recordings in a studio environment help in analyzing the characteristics of the human voice production system responsible for generating speech and singing, without major interference from external factors.

The significant differences between speech and singing lie in prosody (energy, duration, and pitch), and timbre (spectrum)~\cite{KTHSundberg, fujisaki1983dynamic}. We compare each of these characteristics of speech and singing with respect to the NHSS database in the subsequent sections. As duration, pitch, and spectrum of speech and singing vary over different classes of phonemes, we present a phoneme class specific analysis for these three aspects. In the following analysis, all utterances in the NHSS database are downsampled to 16 kHz. 

\vspace{-0.2cm}
\subsection{Energy}\label{sec:energy}
An important characteristic that distinguishes between speech and singing is the variation in loudness, which can be approximated by energy~\cite{seshadri2009perceived,werner1994fundamentals,sharma2019automatic}. In singing, the energy rendered by a singer follows the required melody for a song~\cite{monson2014perceptual,monson2014detection}. In speech, the energy produced by a speaker is generally controlled by the natural speaking manner and emotional state of the speaker. Therefore, the temporal energy contours of parallel utterances of speech and singing signals exhibit significant differences. To study these differences, we compare the dynamic range and distributions of energy between speech and singing. Using the dynamic range of energy we show the extent of vocal loudness expressed by singers as they sing, in comparison with as they speak.
\begin{table}[!h]
 \centering
 \caption{Comparison of dynamic range (DR) of short-time energy (in dB) between parallel utterances of singing and speech in the NHSS database.}
 \renewcommand{\arraystretch}{1.2}
 \vspace{0.2cm}
 \scalebox{0.8}{
 \begin{tabular}{|l|c|c|}
 \hline
 {\bf Statistics of DR (dB)} & {\bf Singing} & {\bf Speech} \\ \hline \hline
 Mean & 33.60 & 30.95  \\ \hline
 Median & 33.64 & 30.56  \\ \hline
 Standard deviation & 3.73 & 3.14  \\ \hline
  \end{tabular}}
 \label{tab:energy}
\end{table}

\begin{figure}[h!]
\vspace{-0.2cm}
\centering
\centerline{\epsfig{figure=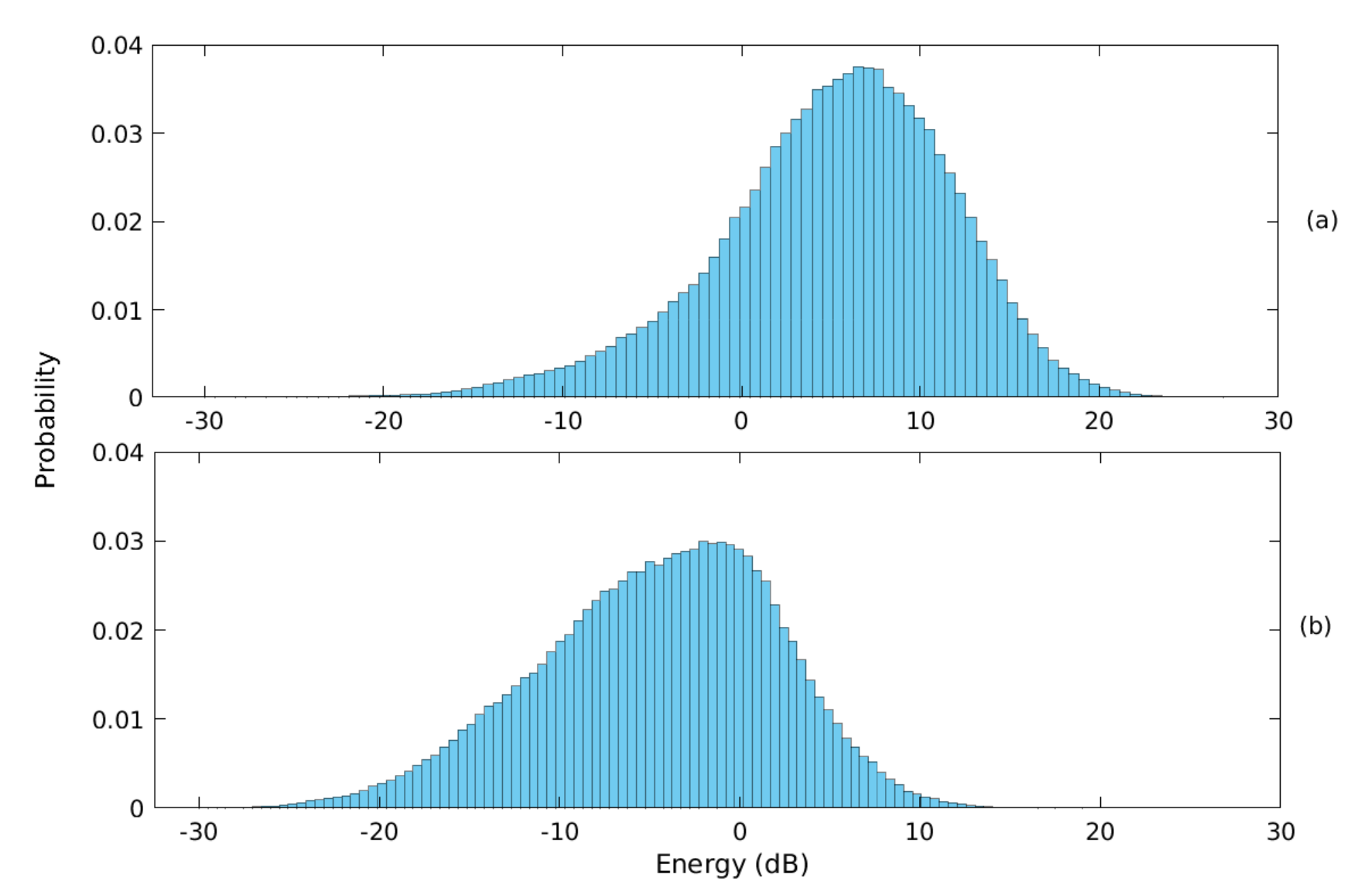,height=2.4 in, width=3.6in}}
\vspace{-0.2cm}
\caption{Histogram distributions of energy for (a) singing, (b) speech for the NHSS database.}
\label{fig:energyhist}
\vspace{-0.1cm}
\end{figure}

To study the temporal variation in energy between speech and singing signals in the NHSS database, we compute short-time energy (in dB) for the signal $x[n]$ as follows. We normalize the signals with respect to their maximum amplitude before calculating the energy. We calculate energy in dB scale to relate to human perception.
\begin{equation}
E[k] = 10\log (\sum_{n=0}^{N-1} x_k^2[n]),
\end{equation}
where, $x_k[n]$ is the $k^{\text{th}}$ frame of the signal, N is framesize (25 msec) and k is the frame index. We consider a frameshift of 25 msec. We normalize the energy contour with respect to the maximum energy across frames and calculate the dynamic range of energy ($DR_E$) as follows.
\begin{equation}
DR_{E}[k] = max(E[k]) - min(E[k]),
\end{equation}
where,  k = 1 to K, and K is the number of frames in an utterance of speech/singing.

In Figure~\ref{fig:energyhist} (a) and Figure~\ref{fig:energyhist} (b), we show the histograms of short-time energy of singing and speech over the entire NHSS database, respectively. We can observe that higher values of short-time energy significantly appear in singing than speech.
In Table~\ref{tab:energy}, we show the dynamic range of short-time energy per utterance and report the mean, median, and standard deviation of dynamic ranges for all speech and singing utterances in the NHSS database. We observe that the average dynamic range of energy for singing is considerably larger than that for speech. Also, the variation of the dynamic range of energy across singing utterances is greater than that of speech. The singing renditions in the NHSS database manifest a large variation in temporal energy as compared to the speech, which is motivated by the melody of songs that are absent while reading lyrics.

\subsection{Duration}

\begin{figure}[h!]
\vspace{-0.2cm}
\centering
\centerline{\epsfig{figure=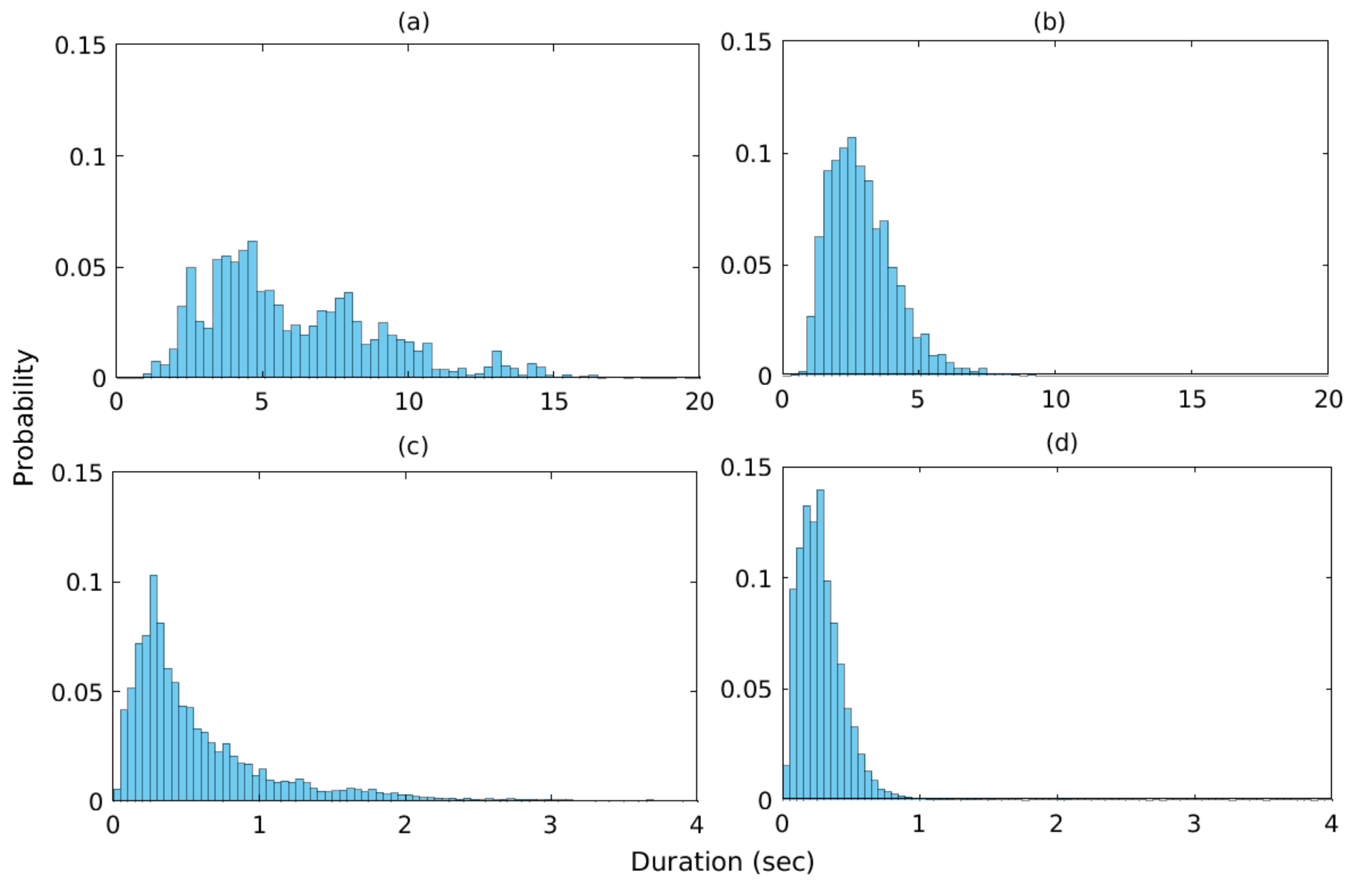,height=2.4 in, width=3.6in}}
\vspace{-0.2cm}
\caption{Histogram distributions of (a) singing utterance duration, (b) speech utterance duration, (c) singing word duration, (d) speech word duration for the NHSS database.}
\label{AllAvgDuration}
\vspace{-0.4cm}
\end{figure}

\begin{table}[!h]
 \centering
 \caption{ Different classes of phonemes used in the comparative analysis of duration, pitch and spectrum for singing and speech.}
 \renewcommand{\arraystretch}{1}
 \vspace{0.2cm}
 \scalebox{0.9}{
 \begin{tabular}{|c|c|}
\hline
\multicolumn{1}{|c|}{\bf Class} & \multicolumn{1}{|c|}{\bf Phonemes}\\
\hline
\multicolumn{1}{|c|}{Vowels} & \multicolumn{1}{|c|}{AA, AE, AH, AO, AW, AY, EH, ER,}\\
\multicolumn{1}{|c|}{} & \multicolumn{1}{|c|}{EY, IH, IY, OW, OY, UH, UW}\\
\hline
\multicolumn{1}{|c|}{Glides} & \multicolumn{1}{|c|}{W, Y}\\
\hline
\multicolumn{1}{|c|}{Liquids} & \multicolumn{1}{|c|}{L, R}\\
\hline
\multicolumn{1}{|c|}{Nasals} & \multicolumn{1}{|c|}{M, N,NG}\\
\hline
\multicolumn{1}{|c|}{Aspirates} & \multicolumn{1}{|c|}{HH}\\
\hline
\multicolumn{1}{|c|}{Fricatives} & \multicolumn{1}{|c|}{DH, S, SH, F, TH, V, Z}\\
\hline
\multicolumn{1}{|c|}{Affricates} & \multicolumn{1}{|c|}{CH, JH}\\
\hline
\multicolumn{1}{|c|}{Stops} & \multicolumn{1}{|c|}{ B , D , G , K , P , T }\\
\hline
\end{tabular}}
\label{tab:PhonemeClasses}
\end{table}

An important difference in the properties of singing over speech is the span of linguistic content with respect to the required melody. The duration of words in the lyrics of a song is stretched or compressed according to the musical notes to be delivered, whereas, in speech the words are pronounced with the normal speaking rate of the speaker. Any kind of analysis involving a comparative study of speech and singing should address the duration variation between the two. In this study, we analyse the utterance, word, and phoneme duration variation between speech and singing.

\begin{table}[!h]
\vspace{-0.4cm}
 \centering
 \caption{Average values of duration of different classes of phonemes across singing and speech for the NHSS database.}
 \renewcommand{\arraystretch}{1}
 \vspace{0.2cm}
 \scalebox{0.9}{
 \begin{tabular}{|c|c|c|}
\hline
\multicolumn{1}{|c|}{\bf Class} & \multicolumn{2}{|c|}{\bf Duration (msec)}\\
\cline{2-3}
\multicolumn{1}{|c|}{} & \multicolumn{1}{|c|}{Singing} & \multicolumn{1}{|c|}{Speech}\\
\hline
\hline
\multicolumn{1}{|c|}{Vowels} & \multicolumn{1}{|c|}{830.5} & \multicolumn{1}{|c|}{148.0}\\
\hline
\multicolumn{1}{|c|}{Glides} & \multicolumn{1}{|c|}{486.8} & \multicolumn{1}{|c|}{102.8}\\
\hline
\multicolumn{1}{|c|}{Liquids} & \multicolumn{1}{|c|}{479.5} & \multicolumn{1}{|c|}{101.4}\\
\hline
\multicolumn{1}{|c|}{Nasals} & \multicolumn{1}{|c|}{586.2} & \multicolumn{1}{|c|}{103.5}\\
\hline
\multicolumn{1}{|c|}{Aspirates} & \multicolumn{1}{|c|}{168.0} & \multicolumn{1}{|c|}{117.7}\\
\hline
\multicolumn{1}{|c|}{Fricatives} & \multicolumn{1}{|c|}{217.5} & \multicolumn{1}{|c|}{116.7}\\
\hline
\multicolumn{1}{|c|}{Affricates} & \multicolumn{1}{|c|}{164.8} & \multicolumn{1}{|c|}{120.5}\\
\hline
\multicolumn{1}{|c|}{Stops} & \multicolumn{1}{|c|}{172.4} & \multicolumn{1}{|c|}{81.5}\\
\hline
\end{tabular}}
\label{tab:DurationPhoneme}
\end{table}

The statistics of the duration of utterances and words over the NHSS database are discussed in Section~\ref{sec:stats}. To further analyse patterns of the duration of the speech and singing audio in the entire NHSS database, we show the histograms of utterance and word duration in Figure~\ref{AllAvgDuration}. We observe from Figure~\ref{AllAvgDuration} (a) and Figure~\ref{AllAvgDuration} (b) that the mean and variance of utterance duration for singing is significantly higher than that of speech. The distributions of word duration shown in Figure~\ref{AllAvgDuration} (c) and Figure~\ref{AllAvgDuration} (d) also indicate that words in singing are pronounced with larger duration variations than singing. 

To study the phoneme specific relative characteristics of duration, we perform phoneme-level automatic alignment of speech and singing over the entire NHSS database. To avoid large errors in phoneme alignment, we consider the manual word labels as anchor points, and apply each word individually to the automatic alignment module. We use MFA~\cite{mcauliffe2017montreal,Montreal} and singing adapted models ~\cite{BidishaICASSP,gupta2020automatic} to obtain the phoneme alignments for speech and singing, respectively. Based on the similarity in production of sounds, we categorize the phonemes into 8 classes as shown in Table~\ref{tab:PhonemeClasses} ~\cite{duan2013nus,sharma2016sonority,sharma2017vowel,sharma2018improving}.

The average phoneme duration for different classes of phonemes is shown in Table~\ref{tab:DurationPhoneme}. We can observe that all the phonemes are stretched in singing than in speech. Specifically, the vowels are observed to have highest stretching. Among the consonants the duration stretching in glides, liquids and nasals is more than other consonants. This shows that in singing the sonorant sounds (vowels, glides, liquids, nasals)
could  be  sustained  and  articulated  for  a  longer time than the non-sonorants (aspirates, fricatives, aspirates, stops).


\subsection{Pitch}
Pitch is another significant property of good quality singing.  Trained singers render pitch transitions in singing by accurately controlling the subglottal pressure and consequently altering the vibration rate of vocal folds at glottis \cite{II-3}. Also, the singers create fine variations in pitch contour, namely, vibrato, overshoot, and preparation, adding to the naturalness of singing voice \cite{fujisaki1983dynamic}. This makes the characteristics of pitch of singing significantly different from speech~\cite{sharma2014faster}.

\begin{table}[!h]
\vspace{-0.4cm}
 \centering
 \caption{Average values of fundamental frequency (F0) of different classes of sounds across singing and speech.}
 \renewcommand{\arraystretch}{1}
 \vspace{0.2cm}
 \scalebox{0.9}{
 \begin{tabular}{|c|c|c|}
\hline
\multicolumn{1}{|c|}{\bf Class} & \multicolumn{2}{|c|}{\bf F0 (Hz)}\\
\cline{2-3}
\multicolumn{1}{|c|}{} & \multicolumn{1}{|c|}{Singing} & \multicolumn{1}{|c|}{Speech}\\
\hline
\hline
\multicolumn{1}{|c|}{Vowels} & \multicolumn{1}{|c|}{268.67} & \multicolumn{1}{|c|}{170.73}\\
\hline
\multicolumn{1}{|c|}{Glides} & \multicolumn{1}{|c|}{207.94} & \multicolumn{1}{|c|}{153.63}\\
\hline
\multicolumn{1}{|c|}{Liquids} & \multicolumn{1}{|c|}{247.99} & \multicolumn{1}{|c|}{159.23}\\
\hline
\multicolumn{1}{|c|}{Nasals} & \multicolumn{1}{|c|}{237.96} & \multicolumn{1}{|c|}{167.11}\\
\hline
\end{tabular}}
\label{tab:PitchPhoneme}
\end{table}

To demonstrate the differences in pitch between speech and singing in the NHSS database, we report the phoneme class specific mean values of fundamental frequency (F0 in Hz) for speech and singing over the entire NHSS database. We  use  the  pitch  estimates  from  the autocorrelation-based  pitch  estimator in Praat~\cite{Praat,rabiner1977use} with 25 msec framesize and 5 msec frameshift. The average pitch values for different classes of phonemes i.e. vowels, glides, liquids and nasals (Table~\ref{tab:PhonemeClasses}) are shown in Table~\ref{tab:PitchPhoneme}. We can observe that the mean F0 of singing is noticeably higher than that of speech for all the classes. Specifically, the vowels are rendered with relatively higher F0 in singing than in speech.

\begin{figure}[h!]
\centering
\centerline{\epsfig{figure=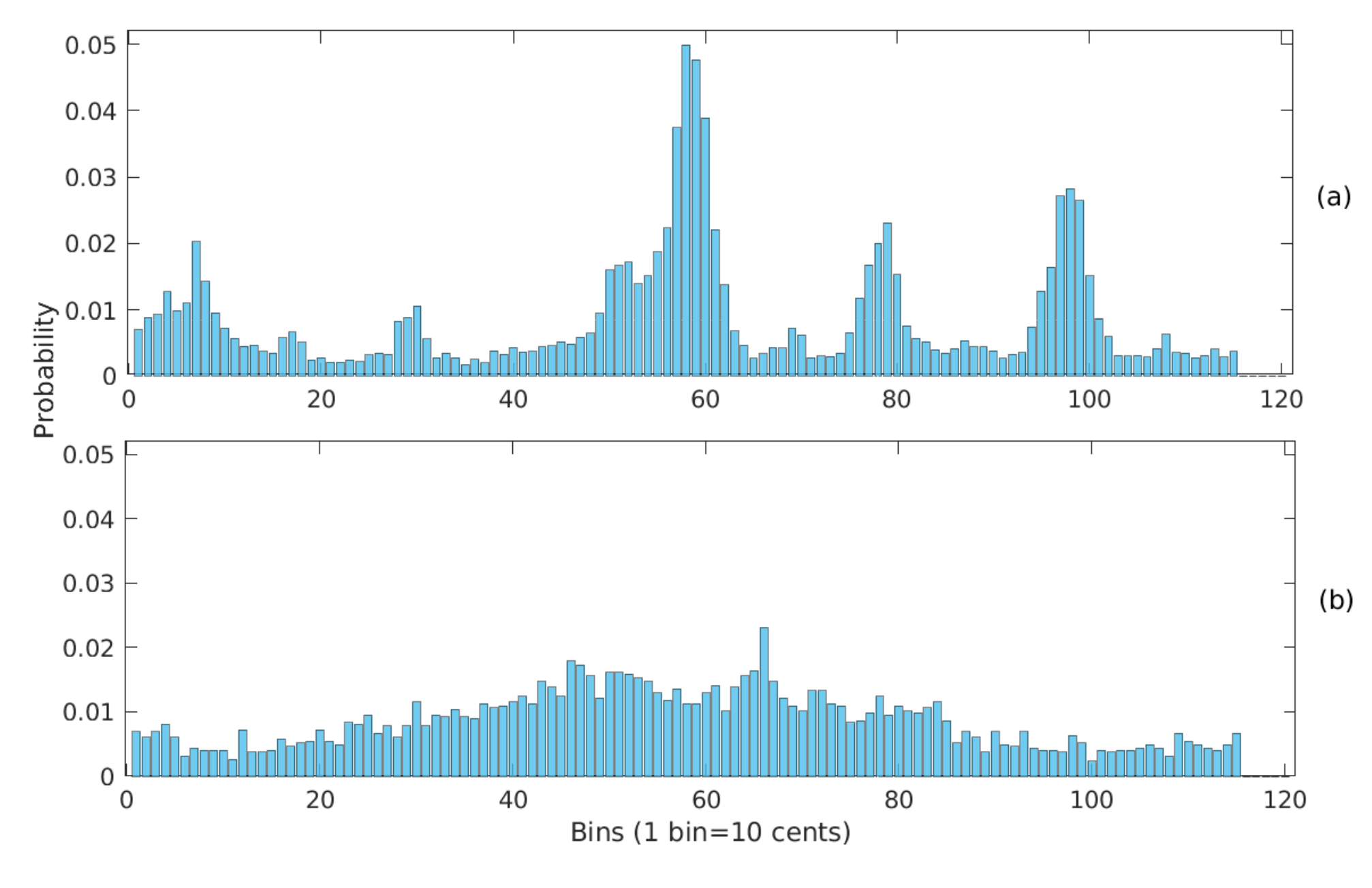,height=2.2 in, width=3.6in}}
\vspace{-0.4cm}
\caption{Pitch histograms of (a) singing, (b) speech corresponding to one song.}
\label{fig:pitchhistogram_1}
\end{figure}

To study the presence of well expressed musical notes in singing, as exhibited by pitch contours, we compare the pitch histograms of speech and singing.
Pitch histogram is a global statistical representation of the pitch content of a musical piece~\cite{tzanetakis2003pitch}. 

To find the pitch histogram we follow the same approach explained in~\cite{Chitra}. We first transform the F0 values from Hz to Cents. To compute the pitch histogram, it is necessary to remove the key of the song.  We subtract the median of pitch values in a singing rendition and transpose all pitch values to a single octave. Then pitch histogram H is computed by placing the pitch values into their corresponding bins.

\begin{equation}
    H_p=\sum_{n=1}^{N} m_p,
\end{equation}
where, $H_p$ is the $p^{\text{th}}$ bin count, N is the number of pitch values. $m_p=1$ if $c_p \leq c_{p+1}$ and $m_p=0$ otherwise, ($c_p,c_{p+1}$) are bounds on $p^{\text{th}}$ bin. We divide each semitone into 10 bins, which results in 120 bins (12 semitones $\times$ 10 bins), each representing 10 cents.

We plot pitch histograms for singing and speech corresponding to one song rendered by a female singer. It can be observed from Figure~\ref{fig:pitchhistogram_1}(a) that the pitch histogram of the singing voice showcases prominent peaks, indicating that pitch values in some ranges are more frequently produced by the singer. This further points to the fact that singers deliver the dominant notes of the song frequently and consistently \cite{Chitra}. On the other hand, the pitch histogram of speech in Figure~\ref{fig:pitchhistogram_1}(b) is relatively flat, showing the absence of specific musical notes.

\begin{figure}[h!]
\centering
\centerline{\epsfig{figure=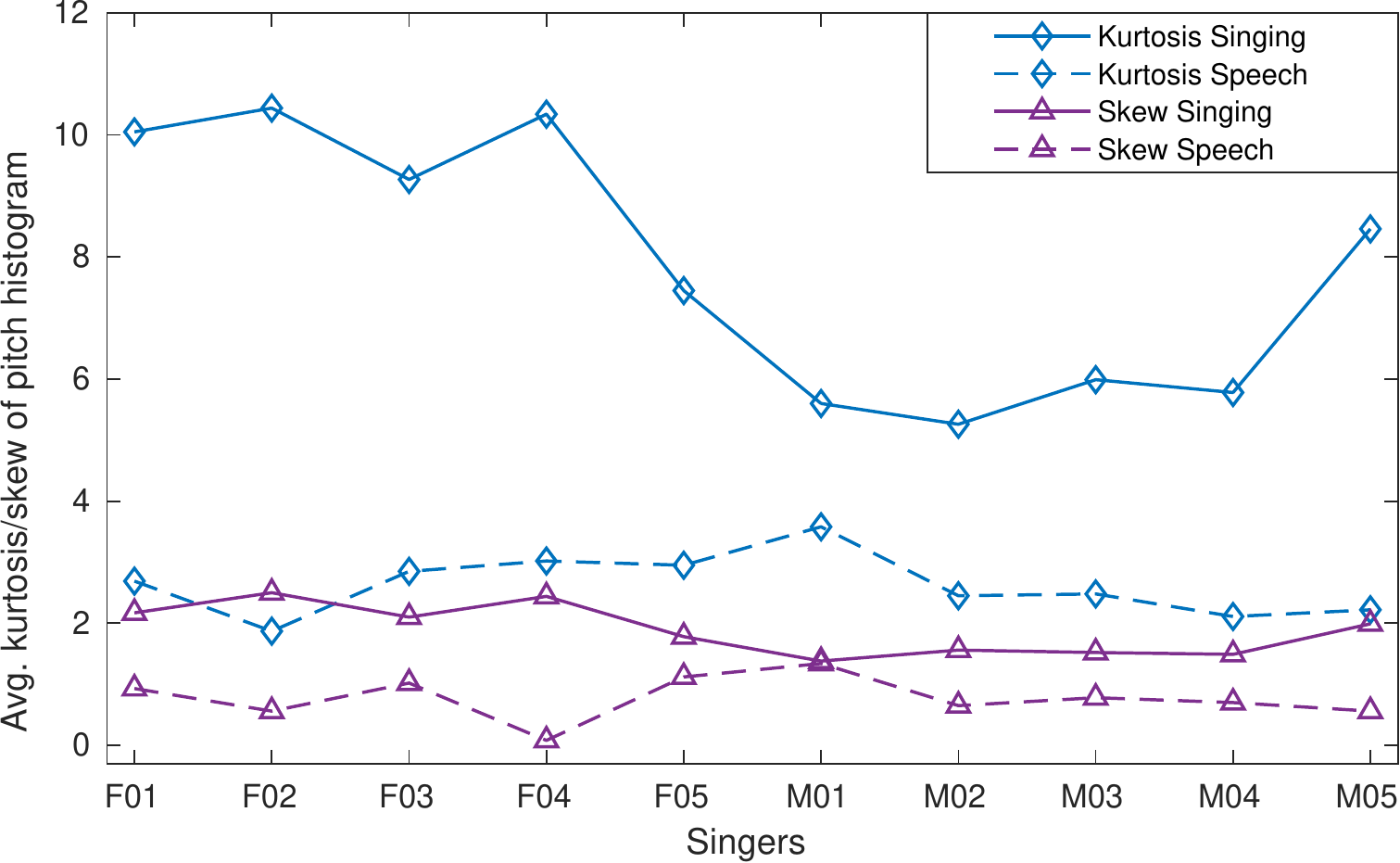,height=2.4 in, width=3.6in}}
\vspace{-0.4cm}
\caption{{Average kurtosis and skew values of pitch histograms for singing and speech corresponding to all songs for each singer.}}
\label{fig:pitchhistogram}
\vspace{-0.5cm}
\end{figure}

To further observe the behavior of pitch histogram across speech and singing over the entire NHSS database, we calculate the {\it kurtosis} and {\it skew} of the pitch histograms~\cite{Chitra} for speech and singing corresponding to each song for all singers. We plot the average kurtosis and skew values over all the songs for each speaker in Figure~\ref{fig:pitchhistogram}. Kurtosis and skew can be used as statistical measures to approximate the sharpness of the pitch histogram, which exhibit higher values for good quality singing~\cite{Chitra}. From Figure~\ref{fig:pitchhistogram}, we can observe that both kurtosis and skew show higher values in singing compared to that of speech for each singer.

A system that is designed for automatic synthesis/conversion of singing voice has to cater to the range of variations, naturalness, and overall quality of pitch renditions.

\subsection{Spectrum}
\label{spectrum}
As the formants in the voice signals are generally decided by the sound being produced, parallel utterances of speech and singing corresponding to the same lyrics should exhibit the same formant frequencies. However, the formant trajectories of singing voice get modulated by the pitch of singing, and hence demonstrate fine variations from respective trajectories in speech. Also in some form of loud singing, like opera, the high frequency spectral region differs between speech and singing due to the presence of singing formant \cite{KTHSundberg}. The loudness variation between speech and singing also contributes to differences in spectral energy between them. To show the spectral differences between speech and singing, we compare the spectral centroid, spectral energy, and first three formants between the two~\cite{grey1978perceptual,sharma2017enhancement}.

The spectral centroid depicts the frequency of a short-time spectrum where its center of mass is located. To calculate the spectral centroid, we first find the spectrum ($S[\omega, k]$) for both speech and singing utterances with 25 msec framesize and 5 msec frameshift using linear prediction coding (LPC) analysis in Praat~\cite{Praat}. The spectral centroid (SC) is calculated as follows.


\begin{equation}
SC[k]= \frac{\sum_{\omega=0}^{\pi}\omega S[\omega, k]}{\sum_{\omega=0}^{\pi}S[\omega, k]},
\end{equation}
where, $\omega$ is the frequency, k=1,2,..K are frame indexes, and K is total number of frames in an utterance. We observe that the spectral centroid of singing is generally higher than that of speech. 

The mean and median values of the spectral centroid of speech and singing over all utterances in the NHSS database are reported in Table~\ref{tab:spectrum}. 
As suggested by the analysis of the spectral centroid, a significant difference in energy content between speech and  singing spectra lies in higher frequency components. To explore this further, we study the spectral energy in the low frequency subband (0 - 2kHz) ($E_l[k]$) and high frequency subband (2 kHz - 4 kHz) ($E_h[k]$). We compute the dynamic range of short-time spectral energy from two subbands over all utterances of speech and singing in the NHSS database, as follows.

\begin{equation}
E_l[k]=\sum_{\omega=0}^{\pi/2} S^2[\omega,k],~~~~E_h[k]=\sum_{\omega=\pi/2}^{\pi} S^2[\omega,k],
\end{equation}
\begin{equation}
DR_{E_l}[k] = 10\log (max(E_l[k]) - min(E_l[k])),
\end{equation}
\begin{equation}
DR_{E_h}[k] = 10\log(max(E_h[k]) - min(E_h[k])),
\end{equation}
where, $DR_{E_l}[k]$ and $DR_{E_h}[k]$ represent the dynamic range of spectral energy corresponding to $k^{\text{th}}$ frame for low and high frequency subbands , respectively. 

From Table~\ref{tab:spectrum}, we observe that the subband spectral energy in singing manifests a larger dynamic range in comparison to that of speech. This observation is particularly more prominent in the higher subband. This large range of variation in higher subband energies between speech and singing may be attributed to the presence of singing formant.
\begin{table}
 \centering
 \caption{Comparison of spectral characteristics (spectral centroid (SC) and dynamic range (DR)) of singing and speech in the NHSS database.}
 \renewcommand{\arraystretch}{1.2}
 \vspace{0.2cm}
 \scalebox{0.8}{
 \begin{tabular}{|l|c|c|}
 \hline
 {\bf Characteristic} & {\bf Singing (Hz)} & {\bf Speech (Hz)} \\ \hline \hline
 Mean of SC  & 967.57 & 843.98\\ \hline
 Median of SC & 618.99 & 507.61  \\ \hline
 DR of spectral energy (0-2 kHz)  & 56.38 & 50.47\\ \hline
 DR of spectral energy (2-4 kHz) & 56.99 & 31.29   \\ \hline
 \end{tabular}}
 \label{tab:spectrum}
 \vspace{-0.3cm}
\end{table}

\begin{table}[!h]
 \centering
 \caption{ Average of first three formant frequencies of different classes of sounds across singing and speech over NHSS database.}
 \renewcommand{\arraystretch}{1.2}
 \vspace{0.2cm}
 \scalebox{0.88}{
 \begin{tabular}{|c|c|c|c|c|c|}
\hline
\multicolumn{6}{|c|}{\bf Formant frequency (Hz)}\\
\hline
\multicolumn{2}{|c}{\backslashbox{\bf Formants$\downarrow$ \kern-3em}{\bf Class$\rightarrow$ }} & \multicolumn{1}{|c|}{\bf Vowels} & \multicolumn{1}{|c|}{\bf Glides} & \multicolumn{1}{|c|}{\bf Liquids} & \multicolumn{1}{|c|}{\bf Nasals}\\
\hline
\hline
\multicolumn{1}{|c}{\multirow{2}{*}{\bf $F_1$}} & \multicolumn{1}{|c}{\bf Singing} & \multicolumn{1}{|c|}{506.84} & \multicolumn{1}{|c|}{320.54} & \multicolumn{1}{|c|}{424.26} & \multicolumn{1}{|c|}{350.01} \\
\cline{2-6}
\multicolumn{1}{|c}{} & \multicolumn{1}{|c}{\bf Speech} & \multicolumn{1}{|c|}{460.37} & \multicolumn{1}{|c|}{293.04} & \multicolumn{1}{|c|}{398.17} & \multicolumn{1}{|c|}{289.67} \\
\hline
\multicolumn{1}{|c}{\multirow{2}{*}{\bf $F_2$}} & \multicolumn{1}{|c}{\bf Singing} & \multicolumn{1}{|c|}{1145.70} & \multicolumn{1}{|c|}{958.84} & \multicolumn{1}{|c|}{1038.50} & \multicolumn{1}{|c|}{1048.10} \\
\cline{2-6}
\multicolumn{1}{|c}{} & \multicolumn{1}{|c}{\bf Speech} & \multicolumn{1}{|c|}{1081.60} & \multicolumn{1}{|c|}{914.57} & \multicolumn{1}{|c|}{949.53} & \multicolumn{1}{|c|}{937.93} \\
\hline
\multicolumn{1}{|c}{\multirow{2}{*}{\bf $F_3$}} & \multicolumn{1}{|c}{\bf Singing} & \multicolumn{1}{|c|}{1931.70} & \multicolumn{1}{|c|}{1740.20} & \multicolumn{1}{|c|}{1716.80} & \multicolumn{1}{|c|}{1871.40} \\
\cline{2-6}
\multicolumn{1}{|c}{} & \multicolumn{1}{|c}{\bf Speech} & \multicolumn{1}{|c|}{1896.20} & \multicolumn{1}{|c|}{1667.90} & \multicolumn{1}{|c|}{1622.70} & \multicolumn{1}{|c|}{1750.60} \\
\hline
\end{tabular}}
\label{tab:FormantPhoneme}
\end{table}
To analyse the variation in formants between speech and singing signals, we report the average formant frequency values for the first ($\text F_1$), second  ($\text F_2$) and third ($\text F_3$) formants in Table~\ref{tab:FormantPhoneme} for singing and speech corresponding to different phoneme classes. The formants frequency values are obtained using LPC analysis in Praat  with 25  msec  framesize  and  5  msec frameshift . We observe that all formants tend to exhibit larger value in singing as compared to their respective speech counterparts.

With a clear understanding of characteristic similarities and dissimilarities between speech and singing, we proceed to present two practically significant case studies using the NHSS database in the subsequent sections.

\section{Speech-to-Singing Alignment}\label{Sec4}
 In this section, we discuss methods for speech-to-singing alignment and provide benchmark results on the NHSS database.

In order to study the relative characteristics between speech and singing and to facilitate conversion from one to another, we often need to establish a frame-level mapping between the two. This mapping between the frames of speech and singing is referred to as speech-to-singing alignment, which can be useful in speech-to-singing conversion. It has been observed that as the temporal alignment degrades, the perceptual quality of speech-to-singing converted samples drastically deteriorates \cite{vijayan2018analysis}. 

Speech-to-singing alignment relies on the invariant information between the two signals. With the precise manual annotations of words in the NHSS database, we are able to study such invariant properties, and evaluate the existing techniques for speech-to-singing alignment.

The dynamic time warping (DTW) technique has been widely used in the literature to temporally align two speech signals~\cite{sakoe1990dynamic}. In the context of speech-to-singing alignment, DTW is performed between the speech and singing signals of the same lyrical content~\cite{cen2012template}. DTW is generally operated over short-time features extracted from the two signals. In this section, we briefly discuss the following three state-of-the-art methods and establish benchmarks for speech-to-singing alignment on the NHSS database. 

\subsection{MFCC Features with DTW}

The MFCCs are widely used features of speech signals for DTW-based temporal alignment with faithful performance. The same strategy of temporal alignment with DTW operating on the MFCC features extracted from speech and singing signals is used for speech-to-singing alignment \cite{cen2012template, cen2011segmentation}. However, the MFCCs may not be the efficient set of features as speech and singing manifest different spectral characteristics~\cite{vijayan2018analysis}. It would be interesting to observe the impact of such disparity reflected in MFCC features.

\subsection{Tandem Features with DTW}
\label{Tandem features with DTW}
There has been a quest for the invariant properties between speech and singing. Based on the early version of NHSS, an idea of tandem features was formulated~\cite{vijayan2018analysis} and evaluated for speech-singing frame alignment. The tandem features include low-time cepstra, voiced-unvoiced decision, and cepstral coefficients from low frequency restricted spectra. The NHSS database facilitates such comparative studies between speech and singing.

\subsection{Model-based Alignment with DTW}

\begin{figure}[h!]
\centering
\centerline{\epsfig{figure=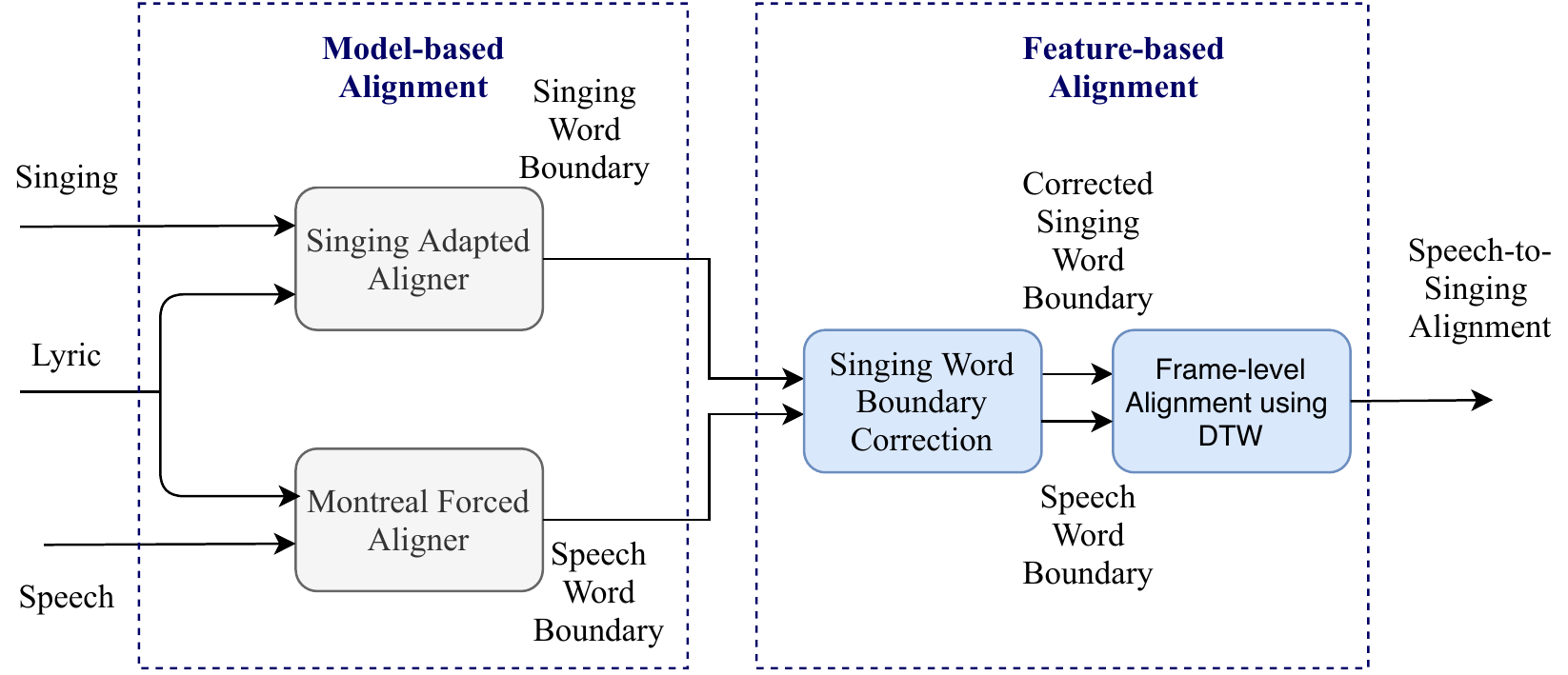,height=1.5 in, width=3.5in}}
\caption{Temporal alignment between speech and singing using model-based alignment with DTW.}
\label{ModelBasedAlignement}
\end{figure}

The DTW technique compares MFCCs or tandem features numerically without referring to any prior knowledge, either spectral, phonetic, or linguistic. We call it the feature-based approach. The model-based approach is a speech recognition based technique that employs acoustic models from speech and singing to perform text-waveform alignment~\cite{sharma2019combination}. As depicted in Figure~\ref{ModelBasedAlignement}, we first obtain the speech and singing word boundaries from model-based alignment, we then apply feature-based DTW alignment to obtain the frame-level alignment.

We note that in this study, we do not use the available manual word boundaries as anchor points, instead we employ acoustic models to initially find the word boundaries. This generalizes the method to be applicable for any pair of parallel speech and singing signals, without any word boundary information.

The model-based forced alignment benefits from the acoustic model that is trained on large database to obtain a coarse alignment at word-level. This is followed by a feature-based DTW to obtain a fine alignment at frame level. The two-stage alignment strategy avoids the propagated error when applying DTW over long utterances. This study is reported in~\cite{sharma2019combination} using an earlier version of the NHSS database. The study suggests that speech-to-singing alignment relies on both adequate acoustic features and effective alignment techniques. Speech-to-singing alignment remains a challenge that is worthy of further research.

\subsection{Experiments}

For evaluation of the three alignment methods discussed above, we consider all speech and singing utterances in the NHSS database. Each of the above mentioned methods is applied to obtain the frame-level temporal alignment between the paired speech and singing utterances.


The alignment error is reported in terms of median word boundary deviation (MWBD). To compute the MWBD, we use the manually marked word labels for speech and singing in the NHSS database as ground-truth timestamps. To calculate MWBD, we first find the aligned frame of singing corresponding to  each manually marked word boundary in speech. The difference between this aligned frame of singing and corresponding manually marked singing word boundary is referred to as word boundary deviation (WBD)~\cite{sharma2019combination,vijayan2018analysis}. MWBD is obtained by calculating the median value of WBD over all the utterances. This speech-to-singing alignment error is significantly different from the word boundary deviation reported in Table~\ref{ManualAlignmentSpeech}.

We report the alignment errors in Table~\ref{tab:align}. We can observe a high MWBD of 0.181 sec with MFCC features. The performance improves to 0.165 with the tandem features. However, the model-based alignment with DTW gives the best performance among the three methods under consideration. We note that feature-based techniques are straightforward that give reasonable frame-level alignment, while the model-based technique is more accurate, that depends on external resources, such as speech recognition engines. The model-based technique typically only provides phoneme-level alignment. A combination of feature-based and model-based methods is therefore a promising solution.


\begin{table}
 \centering
 \caption{Median word boundary deviation (MWBD) resulting from DTW  in speech-to-singing alignment over the entire NHSS database.}
 \renewcommand{\arraystretch}{1.4}
 \vspace{0.2cm}
 \scalebox{0.9}{
 \begin{tabular}{|c|c|c|}
  \hline
  {\bf DTW alignment} & {\bf Feature dim} & {\bf MWBD (sec)} \\ \hline \hline
  MFCC features & 60 & 0.181 \\ \hline
  Tandem features & 45 & 0.165\\ \hline
  Model-based alignment & 39 & 0.072 \\ \hline
 \end{tabular}}
\label{tab:align}
\end{table}

With the recent advances of alignment techniques, temporally aligned parallel utterances can be employed for a multitude of applications, such as analysis of voice production mechanism behind speech and singing sounds, speech-to-singing conversion, and speech/singing synthesis studies.  

\section{Speech-to-Singing Conversion}\label{Sec5}

We carry out another experiment on NHSS for speech-to-singing conversion , that serves as a reference system for readers. In speech-to-singing conversion, we convert the read speech by a user (user's speech) into his/her singing with the same lyrical content. The basic idea of speech-to-singing conversion is to transform the prosody and spectral features from the user's speech to those of the reference singing while preserving the speaker identity of the user. In this work, we particularly focus on template-based speech-to-singing conversion, where we keep the prosody same as that of the reference singing.


As the NHSS database consists of parallel speech and singing data from multiple speakers, it facilitates the study of both speaker dependent and speaker independent spectral mapping/conversion from speech to singing. We perform neural network based spectral mapping to convert the spectral features of speech to that of the singing voice. Although the amount of data in the NHSS database limited, this database is sufficient for training a neural network model for spectral mapping~\cite{BLSTM}. The speech-to-singing conversion methods presented in this section are previously explored in~\cite{gao2019speaker}. The difference between the experiments presented in this work and in~\cite{gao2019speaker} is in terms of the amount of train and test data. In this work, we aim to use the entire  NHSS database for the experiments to provide reference systems, and to demonstrate it's application specific significance.

We compare three speech-to-singing conversion approaches, namely, zero-effort transfer, speaker-dependent spectral mapping, and speaker-independent spectral mapping. In all the methods, we use the manually annotated word boundaries and tandem features with DTW to obtain the alignment between the frames of speech and singing.

\subsection{Zero-effort Transfer}\label{Template-based Speech-to-Singing System}

\begin{figure}[h!]
\centering
\flushleft
\includegraphics[width=90mm]{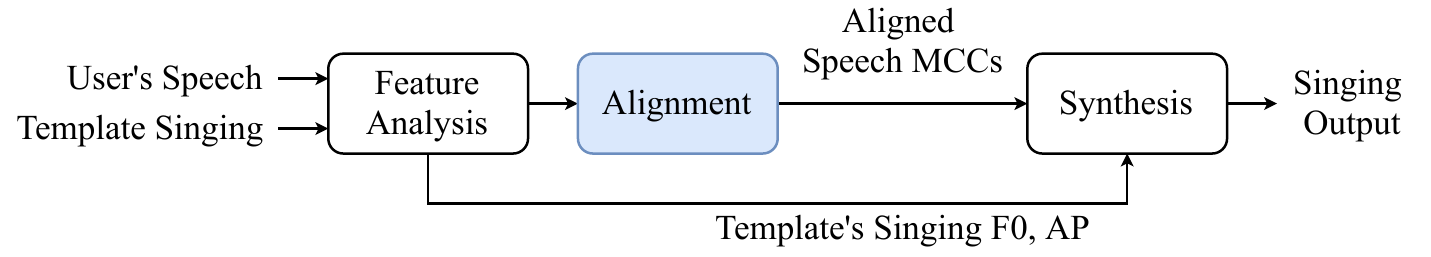}
\vspace{-5mm}
\caption{Block diagram of the zero-effort transfer method.}
\label{fig:TSTS}
\vspace{-5mm}
\end{figure}

Zero-effort transfer refers to the template-based speech-to-singing conversion without spectral mapping. 
Figure~\ref{fig:TSTS} demonstrates the process of the zero-effort transfer approach. Given the user's speech and template singing utterances, we first extract Mel cepstral coefficients (MCCs), F0, and aperiodic component (AP) from both.
Then we employ DTW~\cite{vijayan2018analysis,sakoe1990dynamic} on the tandem features to obtain temporal alignment between the user's speech and template singing, as discussed in Section~\ref{Tandem features with DTW}.
According to the frame-wise temporal alignment information, we align the speech MCCs to match the template singing duration. With aligned speech MCCs, and the F0 and AP of template singing, we generate the converted singing voice through the synthesis module. 


\subsection{Speaker-Dependent Spectral Mapping}
\label{Speaker Dependent Spectral Mapping System}

As discussed in Section~\ref{spectrum}, there exists significant differences between speech and singing spectrum, even for the same linguistic content. To improve the converted singing voice quality, we can use spectral mapping to accommodate singing specific attributes into the speech spectra while retaining the voice identity of the speaker.
Speaker-dependent (SD) spectral mapping model learns the transformation to map speech spectra to singing for each user~\cite{gao2019speaker}.

\begin{figure}[h!]
\centering
\vspace{-0.3cm}
\flushleft
\includegraphics[width=90mm]{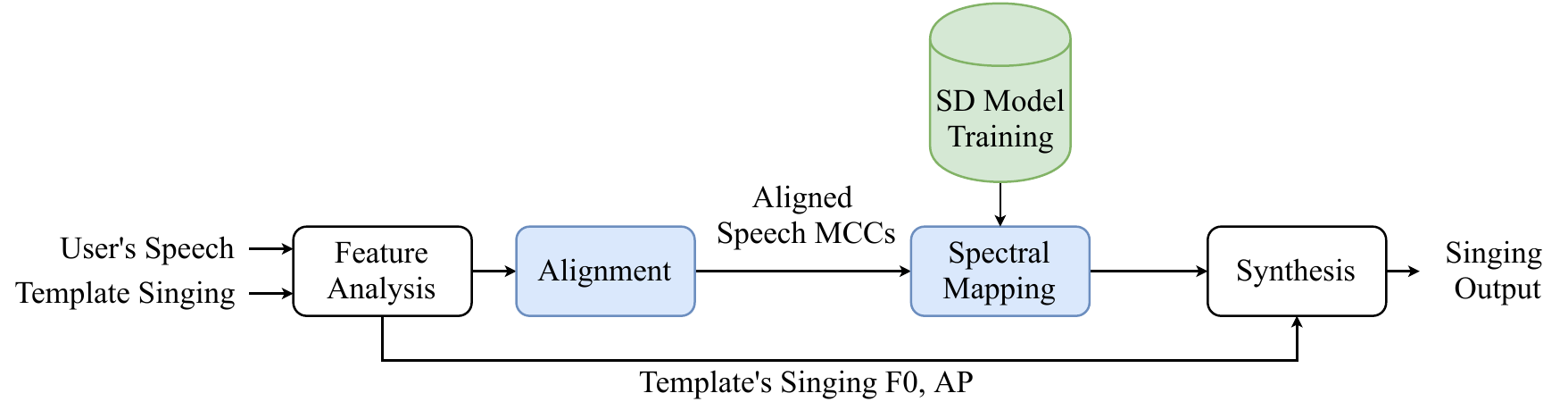}
\vspace{-5mm}
\caption{Block diagram for the run-time conversion stage of the speaker-dependent (SD) spectral mapping. The SD model is trained as described in Section~\ref{Speaker Dependent Spectral Mapping System}.}
\label{fig:SD}
\end{figure}


At the run-time conversion stage shown in Figure~\ref{fig:SD}, given the user's speech and template singing, we first obtain AP and F0 features from template singing and aligned speech MCCs from the user's speech, as discussed in Section~\ref{Template-based Speech-to-Singing System}. The aligned speech MCCs are then taken as the input for the trained SD model to predict the singing MCCs. The converted singing sample is generated using the converted MCCs together with the template's singing F0 and AP parameters.



\subsection{Speaker-Independent Spectral Mapping}\label{Speaker-independent Spectral Mapping System}

\begin{figure}[h!]
\centering
\vspace{-2mm}
\flushleft
\includegraphics[width=90mm]{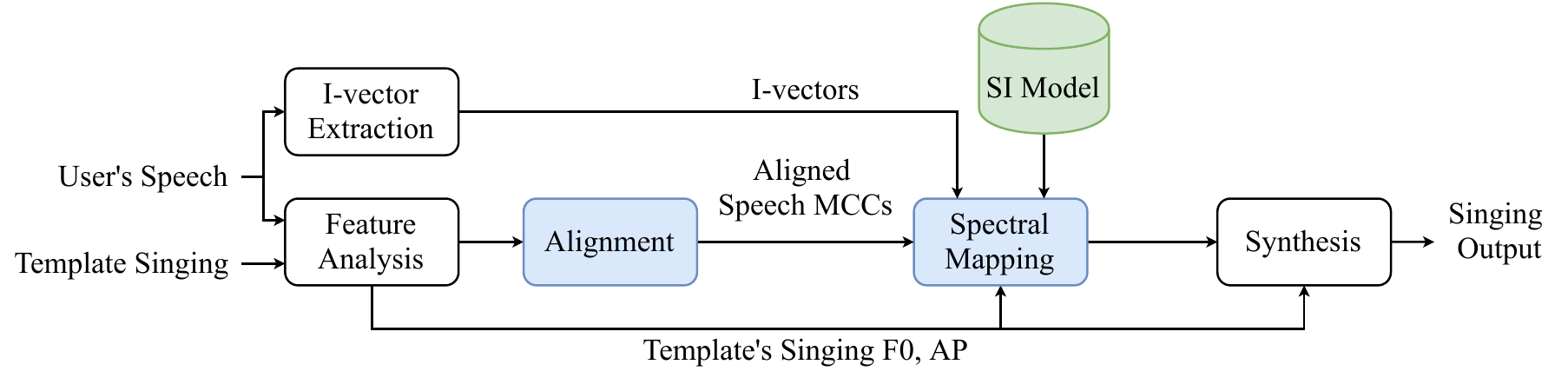}
\vspace{-5mm}
\caption{Block diagram for the run-time conversion stage of the speaker-independent (SI) spectral mapping. The SI model is trained as described in Section~\ref{Speaker-independent Spectral Mapping System}.}
\label{fig:SI}
\vspace{+1mm}
\end{figure}

The parallel speech and singing voice recordings from multiple singers in the NHSS database facilitate to develop speaker-independent (SI) spectral mapping model for speech-to-singing conversion. In SI spectral mapping, we train an average transformation function with parallel speech and singing data from multiple speakers. While an i-vector is used to control the speaker identity of generated singing. During the training stage of SI spectral mapping, we use i-vectors as part of the input features to differentiate speakers during the SI model training~\cite{gao2019speaker}.
Moreover, it is known that the amplitudes of the formants in singing voice are modulated in synchronization with the vibratos in singing F0 contours \cite{new2010voice,henrich2011vocal,oncley1971frequency,fujisaki1983dynamic}. Hence, we introduce the singing F0 and AP parameters as the joint input features with aligned speech MCCs and i-vectors to form the acoustic features to train the SI spectral mapping model.

The run-time conversion is similar to that of the speaker dependent spectral mapping except that the spectral mapping is now conditioned on the speaker's i-vector as shown in Figure~\ref{fig:SI}. The trained SI model is used to map the input features to the converted singing MCCs. The converted singing can be finally synthesized using the converted MCCs and template's singing F0, AP parameters.

\subsection{Experimental Setup}\label{Sec6}
We now compare the three speech-to-singing conversion approaches through experiments on the NHSS database.

\begin{itemize}
    
    \item \textbf{Zero-effort}: The template-based speech-to-singing conversion system without spectral mapping, as described in Section~\ref{Template-based Speech-to-Singing System}.
  
    \item \textbf{SD}: The template-based speech-to-singing conversion system with SD spectral mapping, as described in Section~\ref{Speaker Dependent Spectral Mapping System}, where the input features vector (120-dim) consists of aligned speech MCCs (40-dim) and their delta and delta-delta coefficients.
    \item \textbf{SI}: The template-based speech-to-singing conversion system with SI spectral mapping using i-vectors, as described in Section~\ref{Speaker-independent Spectral Mapping System}. In this case, the input feature (187-dim), includes aligned speech MCCs (40-dim), singing $\log F0$ (1-dim), singing AP (1-dim) with their delta and delta-delta coefficients, the voiced/unvoiced flag (1-dim) and i-vectors (60-dim).
\end{itemize} 

\begin{table} [h!]
\vspace{-0.4cm} 
\caption{\label{table:datades} {Details of the NHSS database in the three speech-to-singing experiments.}}
\renewcommand{\arraystretch}{1.1}
\vspace{.2cm}
\centerline{
\scalebox{0.78}{
\begin{tabular}{|c c c c c|}
\hline
 \multicolumn{5}{|c|}{\bf{Training stage}}\\
 \hline
\multicolumn{1}{|c}{\bf{Method$\rightarrow$}} & \multicolumn{1}{|c|}{\bf{Zero-effort}} & \multicolumn{2}{|c|}{\bf{SD}} & \multicolumn{1}{|c|}{\bf{SI}}\\
\hline
\multicolumn{1}{|c|}{Speakers/singers} & \multicolumn{1}{|c|}{N/A} &
\multicolumn{1}{|c|}{M03} & \multicolumn{1}{|c|}{F03} & \multicolumn{1}{|c|}{All except M03 \& F03}\\
\hline
\multicolumn{1}{|c|}{\#Songs} & \multicolumn{1}{|c|}{N/A} &
\multicolumn{1}{|c|}{8} & \multicolumn{1}{|c|}{8} &\multicolumn{1}{|c|}{80}\\
\hline
\multicolumn{1}{|c|}{\#Utterances} & \multicolumn{1}{|c|}{N/A} &
\multicolumn{1}{|c|}{240} &\multicolumn{1}{|c|}{226} &\multicolumn{1}{|c|}{2,212}\\
\hline
\hline
 \multicolumn{5}{|c|}{\bf{Testing stage}}\\
 \hline
\multicolumn{1}{|c}{\bf{Method$\rightarrow$}} & \multicolumn{4}{|c|}{\bf{Zero-effort/SD/SI}}\\
\hline
\multicolumn{1}{|c|}{Speakers/singers} & \multicolumn{2}{|c|}{M03} &
\multicolumn{2}{|c|}{F03}\\
\hline
\multicolumn{1}{|c|}{\#Songs} & \multicolumn{2}{|c|}{2} &
\multicolumn{2}{|c|}{2}\\
\hline
\multicolumn{1}{|c|}{\#Utterances} & \multicolumn{2}{|c|}{47} &
\multicolumn{2}{|c|}{55}\\
\hline
\end{tabular}}
}
\end{table} 
 
 

\subsubsection{Data Description}
In Table~\ref{table:datades}, we present a summary of the data used for different experiments. We train two SD models on one male and one female singer, namely M03 and F03, with 8 songs of each singer. While an SI model is trained with 8 singers, including 4 male (M01, M02, M04, M05) and 4 female (F01, F02, F04, F05) singers. In total, the training data consists of 2,212 utterances from 80 songs. We use 102 non-overlapping utterances from M03 and F03 for the evaluation of all three models.
All the objective and subjective evaluation results are reported based on the same test set. In this experiment, all the audio signals are downsampled to 16 kHz.

\subsubsection{Feature Extraction}
We use WORLD vocoder \cite{morise2016world} to extract the spectrum (513-dim), AP (1-dim), and F0 (1-dim) for both speech and singing utterances with 25 ms framesize and 5 msec framehift. We compute 40-dimensional MCCs from the spectrum using the speech signal processing toolkit (SPTK)~\footnote{\url{https://sourceforge.net/projects/sptk/}}. The 60-dimensional i-vectors are extracted for each speaker as discussed in ~\cite{dehak2010front}.

\subsubsection{Network Configuration}
We use the same neural network architecture for all the spectral mapping models. The network consists of two bidirectional long short-term memory (BLSTM) layers with 512 hidden units in each layer. The network output is 120-dimensional, including singing MCCs (40-dim) with their delta and delta-delta coefficients. Adam optimizer is used for model training with the learning rate and momentum of 0.002 and 0.9, respectively, and the minibatch size is 10. We use Merlin toolkit \cite{wu2016merlin} for the training of all the models.

In the conversion phase, the maximum likelihood parameter generation (MLPG) algorithm is employed to refine the spectral parameter trajectory \cite{tokuda2000speech}, followed by a spectral enhancement post-filtering in the cepstral domain.



\subsection{Objective Evaluation}
We employ Mel-cepstral distortion (MCD) as the objective measure, which is calculated as follows,
\begin{equation}
\text{MCD}[\text{dB}] = \frac{10}{\text{ln}10}\sqrt{2\sum_{d=1}^{D}(c_{d}-c_{d}^{\text{converted}})^{2}},
\end{equation}
where $c_{d}$ and $c_{d}^{\text{converted}}$ are $d^{\text{th}}$ dimension of singing and the converted MCCs, respectively. \textit{D} denotes the dimension of MCCs. A lower MCD value indicates a smaller distortion.


\begin{table} [h!]
\caption{\label{ObjectiveEval} {Objective and subjective evaluation results in terms of Mel-cepstral distortion (MCD) and mean opinion score (MOS) with 95\% confidence intervals, respectively,  for the original singing and the synthesized singing outputs from SI, SD, and zero-effort systems.}}
\renewcommand{\arraystretch}{1}
\vspace{.3cm}
\centerline{
\scalebox{1}{
\begin{tabular}{|c c c|}
\hline
\multicolumn{1}{|c|}{\multirow{2}{*}{\bf Method $\downarrow$}} & \multicolumn{1}{|c|}{\bf{Average MCD}}  & \multicolumn{1}{|c|}{\multirow{2}{*}{\bf MOS}}\\ 
\multicolumn{1}{|c|}{} & \multicolumn{1}{|c|}{\bf{(dB)}} &  \multicolumn{1}{|c|}{}\\ 
\hline
\hline
\multicolumn{1}{|c|}{Zero-effort} &\multicolumn{1}{|c|}{8.89}  &\multicolumn{1}{|c|}{2.80}\\
\hline
\multicolumn{1}{|c|}{SD} & \multicolumn{1}{|c|}{7.28}  &\multicolumn{1}{|c|}{2.92}\\
\hline
\multicolumn{1}{|c|}{SI} &\multicolumn{1}{|c|}{6.85}  &\multicolumn{1}{|c|}{3.55}\\
\hline
\multicolumn{1}{|c|}{Original singing} &\multicolumn{1}{|c|}{N/A}  &\multicolumn{1}{|c|}{4.53}\\
\hline
\end{tabular}}
}
\end{table} 
We present the objective evaluation results in Table~\ref{ObjectiveEval}. From Table~\ref{ObjectiveEval}, we observe that the systems with spectral mapping (SD, SI) perform better with lower MCD values than the system without spectral mapping (zero-effort) (8.89 dB). This suggests that the spectral mapping improves the speech-to-singing conversion. While comparing the performance of SD and SI, we observe that the SI model outperforms the SD model, with average MCD values of 6.85 dB and 7.28 dB, respectively. These observations indicate the effectiveness of the SI model utilizing multi-speaker speech and singing parallel data for speech-to-singing conversion. 

\subsection{Subjective Evaluation}

To further evaluate the quality and naturalness of the converted singing voice, we  conduct three different subjective evaluations, which are mean opinion score (MOS), AB, and XAB preference tests. 
25 listeners participate in all the three tests. For each test, 15 samples are randomly selected from each system, which results in 45 samples for individual listening test. In the MOS test, listeners are asked to rate the quality and naturalness ranging from 1 (poor) to 5 (excellent) for the samples of all three systems. In the AB preference test, one singing sample is picked from system A and another is from system B. A listener is asked to select the better sample in terms of the quality of synthesized singing. In the XAB preference test, given reference target singing X, listeners are asked to choose which one is more similar to X from samples of the systems A and B, to evaluate the converted singing in terms of speaker similarity.  We have provided the converted singing samples corresponding to different speech-to-singing conversion methods for listening in~\footnote{\url{https://hltnus.github.io/NHSSDatabase/demo.html}}


\leftmargini=5mm
\begin{itemize}

\item[\emph{1)}] \emph{Mean opinion score (MOS) test:}
In Table~\ref{ObjectiveEval}, we depict the MOS values of different speech-to-singing conversion systems with 95\% confidence intervals. From Table~\ref{ObjectiveEval}, we can observe that the SI model (3.77) significantly outperforms the SD model (2.94) and zero-effort (2.86) in terms of the naturalness of synthesized/converted singing voice. The better performance of the system with the SI model is expected due to the use of multi-speaker data for training the spectral conversion model. These results also resonate with the objective evaluation performance shown in Table~\ref{ObjectiveEval}. 

\begin{figure}[h!]
\centering
\vspace{-1mm}
\includegraphics[height=36mm, width=85mm]{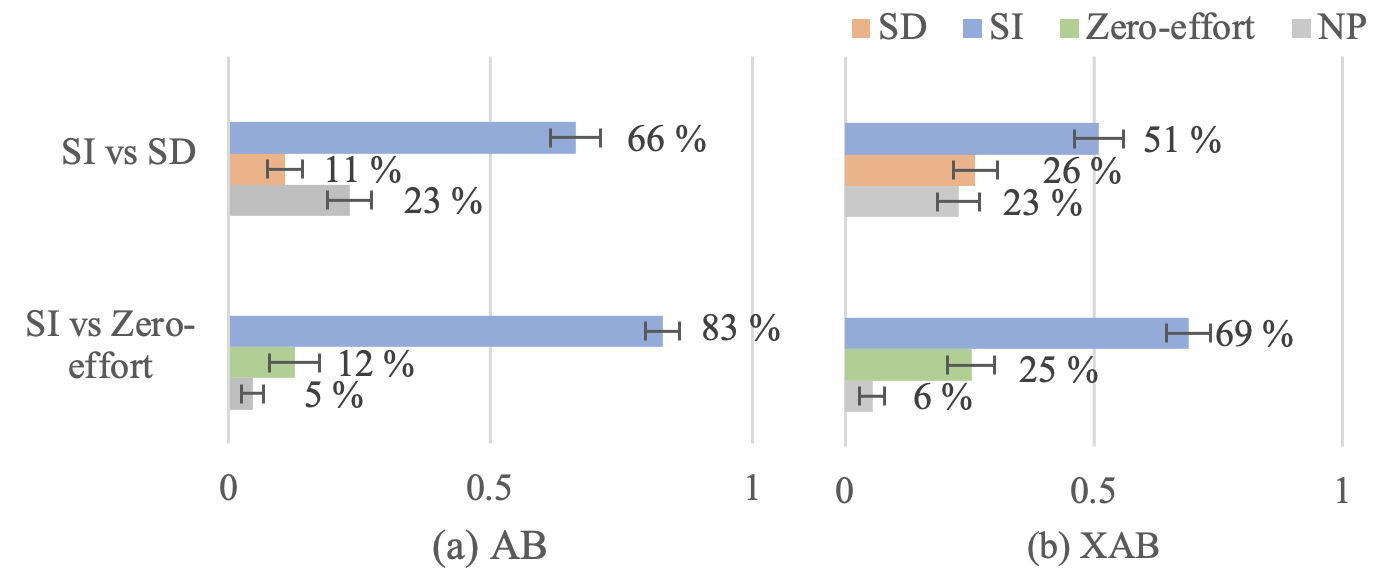}
\vspace{-0.3cm}
\caption{(a) AB preference, and (b) XAB preference test results with 95\% confidence intervals comparing the SI vs SD systems, and SI vs zero-effort systems; NP stands for no preference.}
\label{fig:TSTSvsSI}
\vspace{-1mm}
\end{figure}

\item[\emph{2)}] \emph{AB preference test:}
We carry out an AB preference test to assess the comparative singing quality among different speech-to-singing conversion systems. In Figure~\ref{fig:TSTSvsSI} (a), we present the preference scores with 95\% confidence intervals. We observe that the SI model (89$\%$) significantly outperforms zero-effort (7$\%$), which demonstrates the effectiveness of the spectral mapping in terms of improving the converted singing quality. We also observe that the SI model (81$\%$) performs significantly better than the SD model (6$\%$). This suggests that the SI model benefits from the training on the multi-speaker NHSS database.

\item[\emph{3)}] \emph{XAB preference test:}
We conduct an XAB preference test to assess the speaker similarity for each of the three speech-to-singing conversion systems. In Figure~\ref{fig:TSTSvsSI} (b), we show the preference scores with 95\% confidence intervals.  The performance of the three systems follows the same trend as that of the AB preference test.
Again, we observe that the SI model significantly outperforms both the zero-effort and SD models in terms of speaker similarity.
This suggests that the use of the i-vectors in the SI model are beneficial to retain the target speaker's identity.
\end{itemize}

To summarize, both objective and subjective evaluation results confirm that the SI model trained with multi-speaker data performs significantly better than the zero-effort and SD models. This study demonstrates the prevalent significance of the presented speech and singing parallel database in the development of a speech-to-singing conversion system. 

\vspace{-0.3cm}
\section{Conclusion}\label{Sec8}
In this paper, we present the development of the NHSS database, which consists of 7 hours of speech and singing parallel recordings from multiple singers, along with utterance- and word-level annotations for the entire database. We discuss data collection, processing, and organization of the database in detail. We analyse some similar and distinctive prosodic characteristics of speech and singing with respect to the presented database.

We demonstrate that the NHSS database can be potentially used in the research of speech-to-singing alignment, spectral mapping, and conversion. We carry out two case studies, temporal alignment between speech and singing, and speech-to-singing voice conversion using spectral mapping. The presented systems can be used as the reference systems on the NHSS database. 



The NHSS database can be used to further the comparative study between speech and singing, and benefit applications like lyrics recognition, singer recognition, singing voice conversion, and so on. This database is made publicly available through the website~\footnotemark[1] for research purposes. 

\section{Acknowledgement}
This research work is supported by Academic Research Council, Ministry of Education (ARC, MOE). Grant: MOE2018-T2-2-127. Title: Learning Generative and Parameterized Interactive Sequence Models with RNNs.
{
\bibliographystyle{elsarticle-num}

\bibliography{SpeechSinging.bib}

\begin{thebibliography}{10}
\expandafter\ifx\csname url\endcsname\relax
  \def\url#1{\texttt{#1}}\fi
\expandafter\ifx\csname urlprefix\endcsname\relax\def\urlprefix{URL }\fi
\expandafter\ifx\csname href\endcsname\relax
  \def\href#1#2{#2} \def\path#1{#1}\fi

\bibitem{garofolo1993timit}
J.~S. Garofolo, L.~F. Lamel, W.~M. Fisher, J.~G. Fiscus, D.~S. Pallett,
  N.~LDahlgren, V.~Zue, {TIMIT} acoustic phonetic continuous speech corpus,
  Linguistic Data Consortium (1993).

\bibitem{godfrey1992switchboard}
J.~J. Godfrey, E.~C. Holliman, J.~McDaniel, {SWITCHBOARD}: {Telephone} speech
  corpus for research and development, in: International Conference on
  Acoustics, Speech, and Signal Processing (ICASSP), Vol.~1, 1992, pp.
  517--520.

\bibitem{panayotov2015librispeech}
V.~Panayotov, G.~Chen, D.~Povey, S.~Khudanpur, Librispeech: an {ASR} corpus
  based on public domain audio books, in: IEEE International Conference on
  Acoustics, Speech and Signal Processing (ICASSP), 2015, pp. 5206--5210.

\bibitem{paul1992design}
D.~B. Paul, J.~M. Baker, The design for the {Wall} {Street} {Journal}-based
  {CSR} corpus, in: Proceedings of the workshop on Speech and Natural Language,
  Association for Computational Linguistics, 1992, pp. 357--362.

\bibitem{carletta2005ami}
J.~Carletta, S.~Ashby, S.~Bourban, M.~Flynn, M.~Guillemot, T.~Hain, J.~Kadlec,
  V.~Karaiskos, W.~Kraaij, M.~Kronenthal, et~al., The {AMI} meeting corpus: A
  pre-announcement, in: International workshop on machine learning for
  multimodal interaction, Springer, 2005, pp. 28--39.

\bibitem{hirsch2000aurora}
H.-G. Hirsch, D.~Pearce, The {Aurora} experimental framework for the
  performance evaluation of speech recognition systems under noisy conditions,
  in: ASR2000-Automatic Speech Recognition: Challenges for the new Millenium
  ISCA Tutorial and Research Workshop (ITRW), 2000.

\bibitem{leonard1984database}
R.~Leonard, A database for speaker-independent digit recognition, in:
  International Conference on Acoustics, Speech, and Signal Processing
  (ICASSP), Vol.~9, IEEE, 1984, pp. 328--331.

\bibitem{martin2009nist}
A.~F. Martin, C.~S. Greenberg, {NIST} 2008 speaker recognition evaluation:
  {Performance} across telephone and room microphone channels, in: INTERSPEECH,
  2009, pp. 2579--2582.

\bibitem{Nagrani19}
A.~Nagrani, J.~S. Chung, W.~Xie, A.~Zisserman, Voxceleb: Large-scale speaker
  verification in the wild, Computer Speech \& Language 60 (2020) 101027.

\bibitem{lee2015reddots}
K.~A. Lee, A.~Larcher, G.~Wang, P.~Kenny, N.~Br{\"u}mmer, D.~v. Leeuwen,
  H.~Aronowitz, M.~Kockmann, C.~Vaquero, B.~Ma, et~al., The reddots data
  collection for speaker recognition, in: INTERSPEECH, 2015, pp. 2996--3000.

\bibitem{larcher2014text}
A.~Larcher, K.~A. Lee, B.~Ma, H.~Li, Text-dependent speaker verification:
  {Classifiers}, databases and {RSR2015}, Speech Communication 60 (2014)
  56--77.

\bibitem{kominek2004cmu}
J.~Kominek, A.~W. Black, The {CMU} {Arctic} speech databases, in: Fifth ISCA
  workshop on speech synthesis (SSW5), 2004, pp. 223--224.

\bibitem{sharma2016speech}
B.~Sharma, S.~M. Prasanna, Speech synthesis in noisy environment by enhancing
  strength of excitation and formant prominence., in: INTERSPEECH, 2016, pp.
  131--135.

\bibitem{VCTK}
C.~Veaux, J.~Yamagishi, K.~MacDonald, {CSTR} {VCTK} corpus: {English}
  multi-speaker corpus for {CSTR} voice cloning toolkit, University of
  Edinburgh. The Centre for Speech Technology Research (CSTR)\href
  {http://dx.doi.org/http://dx.doi.org/10.7488/ds/1994}
  {\path{doi:http://dx.doi.org/10.7488/ds/1994}}.

\bibitem{ljspeech17}
K.~Ito, The {LJ} speech dataset, {Available} at,
  \url{https://keithito.com/LJ-Speech-Dataset/} (2017).

\bibitem{King2011TheBC}
S.~J. King, V.~Karaiskos, The blizzard challenge 2011, 2011.

\bibitem{King2013TheBC}
S.~J. King, V.~Karaiskos, The blizzard challenge 2013, 2013.

\bibitem{Meseguer-Brocal_2018}
G.~Meseguer-Brocal, A.~Cohen-Hadria, G.~Peeters, {DALI}: {A} large {Dataset} of
  synchronized {Audio}, {LyrIcs} and notes, automatically created using
  teacher-student machine learning paradigm, in: International Society for
  Music Information Retrieval Conference (ISMIR), 2018, pp. 431--437.

\bibitem{DAMPData}
S.~Sing!, Smule digital archive mobile performances(damp),
  \url{https://ccrma.stanford.edu/damp/}, [Online; accessed 08-January-2020]
  (2010).

\bibitem{gupta2018semi}
C.~Gupta, R.~Tong, H.~Li, Y.~Wang, Semi-supervised lyrics and solo-singing
  alignment, in: International Society for Music Information Retrieval
  Conference (ISMIR), 2018, pp. 600--607.

\bibitem{chan2015vocal}
T.-S. Chan, T.-C. Yeh, Z.-C. Fan, H.-W. Chen, L.~Su, Y.-H. Yang, R.~Jang, Vocal
  activity informed singing voice separation with the {iKala} dataset, in: IEEE
  International Conference on Acoustics, Speech and Signal Processing (ICASSP),
  2015, pp. 718--722.

\bibitem{musdb18}
Z.~Rafii, A.~Liutkus, F.-R. St{\"o}ter, S.~I. Mimilakis, R.~Bittner,
  \href{https://doi.org/10.5281/zenodo.1117372}{The {MUSDB18} corpus for music
  separation} (Dec. 2017).
\newline\urlprefix\url{https://doi.org/10.5281/zenodo.1117372}

\bibitem{bertin2011million}
T.~Bertin-Mahieux, D.~P. Ellis, B.~Whitman, P.~Lamere, The million song
  dataset, in: International Society for Music Information Retrieval Conference
  (ISMIR), 2011, pp. 591--596.

\bibitem{bogdanov2019acousticbrainz}
D.~Bogdanov, A.~Porter, H.~Schreiber, J.~Urbano, S.~Oramas, The
  {AcousticBrainz} genre dataset: Multi-source, multi-level, multi-label, and
  large-scale, in: International Society for Music Information Retrieval
  Conference (ISMIR), 2019, pp. 360--367.

\bibitem{schedl2006investigating}
M.~Schedl, P.~Knees, G.~Widmer, Investigating web-based approaches to revealing
  prototypical music artists in genre taxonomies, in: 1st International
  Conference on Digital Information Management, 2006, pp. 519--524.

\bibitem{mckay2006large}
C.~McKay, D.~McEnnis, I.~Fujinaga, A large publicly accessible prototype audio
  database for music research, in: International Society for Music Information
  Retrieval Conference (ISMIR), 2006, pp. 160--163.

\bibitem{ellis2007classifying}
D.~P. Ellis, Classifying music audio with timbral and chroma features, in:
  International Society for Music Information Retrieval Conference (ISMIR),
  2007, pp. 339--340.

\bibitem{knees2004artist}
P.~Knees, E.~Pampalk, G.~Widmer, Artist classification with web-based data, in:
  International Society for Music Information Retrieval Conference (ISMIR),
  2004, pp. 517--524.

\bibitem{Sharma2019}
B.~Sharma, R.~K. Das, H.~Li, {On the Importance of Audio-Source Separation for
  Singer Identification in Polyphonic Music}, in: INTERSPEECH, 2019, pp.
  2020--2024.
\newblock \href {http://dx.doi.org/10.21437/Interspeech.2019-1925}
  {\path{doi:10.21437/Interspeech.2019-1925}}.

\bibitem{aljanaki2016studying}
A.~Aljanaki, F.~Wiering, R.~C. Veltkamp, Studying emotion induced by music
  through a crowdsourcing game, Information Processing \& Management 52~(1)
  (2016) 115--128.

\bibitem{soleymani20131000}
M.~Soleymani, M.~N. Caro, E.~M. Schmidt, C.-Y. Sha, Y.-H. Yang, 1000 songs for
  emotional analysis of music, in: Proceedings of the 2nd ACM international
  workshop on Crowdsourcing for multimedia, 2013, pp. 1--6.

\bibitem{aljanaki2017developing}
A.~Aljanaki, Y.-H. Yang, M.~Soleymani, Developing a benchmark for emotional
  analysis of music, PloS one 12~(3) (2017) 1--22.

\bibitem{duan2013nus}
Z.~Duan, H.~Fang, B.~Li, K.~C. Sim, Y.~Wang, The {NUS} sung and spoken lyrics
  corpus: {A} quantitative comparison of singing and speech, in: Asia-Pacific
  Signal and Information Processing Association Annual Summit and Conference,
  IEEE, 2013, pp. 1--9.

\bibitem{hsu2009improvement}
C.~L. Hsu, J.~S.~R. Jang, On the improvement of singing voice separation for
  monaural recordings using the {MIR-1K} dataset, IEEE Transactions on Audio,
  Speech, and Language Processing 18~(2) (2009) 310--319.

\bibitem{shi2019addressing}
Y.~Shi, J.~Zhou, Y.~Long, Y.~Li, H.~Mao, Addressing text-dependent speaker
  verification using singing speech, Applied Sciences 9~(13) (2019) 2636.

\bibitem{shi_2019_3241566}
Y.~Shi, J.~Zhou, Y.~Long, Y.~Li, H.~Mao,
  \href{https://doi.org/10.5281/zenodo.3241566}{Rss database} (Jun. 2019).
\newblock \href {http://dx.doi.org/10.5281/zenodo.3241566}
  {\path{doi:10.5281/zenodo.3241566}}.
\newline\urlprefix\url{https://doi.org/10.5281/zenodo.3241566}

\bibitem{gao2018nus}
X.~Gao, B.~Sisman, R.~K. Das, K.~Vijayan, {NUS-HLT} spoken lyrics and singing
  ({SLS}) corpus, in: International Conference on Orange Technologies (ICOT),
  2018, pp. 1--6.

\bibitem{II-3}
B.~Lindblom, J.~Sundberg, The human voice in speech and singing, in: Springer
  Handbook of Acoustics, Springer, New York, 2014, pp. 703--746.

\bibitem{fujisaki1983dynamic}
H.~Fujisaki, Dynamic characteristics of voice fundamental frequency in speech
  and singing, in: The production of speech, Springer, 1983, pp. 39--55.

\bibitem{KTHSundberg}
J.~Sundberg, The level of the `singing formant' and the source spectra of
  professional bass singers, Quarterly Progress and Status Report: STL-QPSR
  11~(4) (1970) 21--39.

\bibitem{vijayan2018speech}
K.~Vijayan, H.~Li, T.~Toda, Speech-to-singing voice conversion: The challenges
  and strategies for improving vocal conversion processes, IEEE Signal
  Processing Magazine 36~(1) (2018) 95--102.

\bibitem{dong2014i2r}
M.~Dong, S.~W. Lee, H.~Li, P.~Chan, X.~Peng, J.~W. Ehnes, D.~Huang, {I2R}
  speech2singing perfects everyone's singing, in: INTERSPEECH, 2014, pp.
  2148--2149.

\bibitem{cen2012template}
L.~Cen, M.~Dong, P.~Chan, Template-based personalized singing voice synthesis,
  in: International Conference on Acoustics, Speech and Signal Processing
  (ICASSP), IEEE, 2012, pp. 4509--4512.

\bibitem{Gupta2019NUSS}
C.~Gupta, K.~Vijayan, B.~Sharma, X.~Gao, H.~Li, {NUS} speak-to-sing : A web
  platform for personalized speech-to-singing conversion, in: Show \& Tell,
  INTERSPEECH, 2019, pp. 2376--2377.

\bibitem{gao2019speaker}
X.~Gao, X.~Tian, R.~K. Das, Y.~Zhou, H.~Li, Speaker-independent spectral
  mapping for speech-to-singing conversion, in: IEEE Asia-Pacific Signal and
  Information Processing Association Annual Summit and Conference (APSIPA ASC),
  2019, pp. 159--164.

\bibitem{sundberg1990science}
J.~Sundberg, T.~D. Rossing, The science of singing voice, Vol.~87, ASA, 1990.

\bibitem{dayme2009dynamics}
M.~A. Dayme, Dynamics of the singing voice, Springer Science \& Business Media,
  2009.

\bibitem{saitou2004analysis}
T.~Saitou, N.~Tsuji, M.~Unoki, M.~Akagi, Analysis of acoustic features
  affecting ``singing-ness" and its application to singing-voice synthesis from
  speaking-voice, in: INTERSPEECH, 2004, pp. 1925--1928.

\bibitem{ohishi2006human}
Y.~Ohishi, M.~Goto, K.~Itou, K.~Takeda, On the human capability and acoustic
  cues for discriminating the singing and the speaking voices, 2006, pp.
  1831--1837.

\bibitem{mcauliffe2017montreal}
M.~McAuliffe, M.~Socolof, S.~Mihuc, M.~Wagner, M.~Sonderegger, Montreal forced
  aligner: Trainable text-speech alignment using kaldi., in: INTERSPEECH, 2017,
  pp. 498--502.

\bibitem{Montreal}
Montreal-forced-aligner,
  \url{https://github.com/MontrealCorpusTools/Montreal-Forced-Aligner},
  [Online; accessed 26-Nobvember-2018].

\bibitem{povey2011kaldi}
D.~Povey, A.~Ghoshal, G.~Boulianne, L.~Burget, O.~Glembek, N.~Goel,
  M.~Hannemann, P.~Motlicek, Y.~Qian, P.~Schwarz, et~al., The {Kaldi} speech
  recognition toolkit, in: IEEE workshop on automatic speech recognition and
  understanding, no. EPFL-CONF-192584, IEEE Signal Processing Society, 2011.

\bibitem{mesaros2010automatic}
A.~Mesaros, T.~Virtanen, Automatic recognition of lyrics in singing, EURASIP
  Journal on Audio, Speech, and Music Processing (2010) ArticleID : 546047.

\bibitem{BidishaICASSP}
B.~Sharma, C.~Gupta, H.~Li, Y.~Wang, Automatic lyrics-to-audio alignment on
  polyphonic music using singing-adapted acoustic models, in: International
  Conference on Acoustics, Speech and Signal Processing (ICASSP), IEEE, 2019,
  pp. 396--400.

\bibitem{Praat}
P.~Boersma, D.~Weenink, Praat: doing phonetics by computer,
  \url{http://www.fon.hum.uva.nl/praat/}, [Online; accessed 08-April-2020].

\bibitem{I-5}
J.~Sundberg, I.~R. Titze, R.~Scherer, Phonatory control in male singing: A
  study of the effects of subglottal pressure, fundamental frequency, and mode
  of phonation on the voice source, Journal of Voice 7~(1) (1993) 15 -- 29.

\bibitem{I-69}
C.~T. Herbst, M.~Hess, F.~Müller, J.~G. Švec, J.~Sundberg, Glottal adduction
  and subglottal pressure in singing, Journal of Voice 29~(4) (2015) 391 --
  402.

\bibitem{seshadri2009perceived}
G.~Seshadri, B.~Yegnanarayana, Perceived loudness of speech based on the
  characteristics of glottal excitation source, The Journal of the Acoustical
  Society of America 126~(4) (2009) 2061--2071.

\bibitem{werner1994fundamentals}
S.~Werner, E.~Keller, Fundamentals of speech synthesis and speech recognition:
  Basic concepts, state of the art, and future challenges, chapter prosodic
  aspects of speech, E. Keller (1994) 23--40.

\bibitem{sharma2019automatic}
B.~Sharma, Y.~Wang, Automatic evaluation of song intelligibility using singing
  adapted stoi and vocal-specific features, IEEE/ACM Transactions on Audio,
  Speech, and Language Processing 28 (2019) 319--331.

\bibitem{monson2014perceptual}
B.~B. Monson, E.~J. Hunter, A.~J. Lotto, B.~H. Story, The perceptual
  significance of high-frequency energy in the human voice, Frontiers in
  psychology 5 (2014) 587.

\bibitem{monson2014detection}
B.~B. Monson, A.~J. Lotto, B.~H. Story, Detection of high-frequency energy
  level changes in speech and singing, The Journal of the Acoustical Society of
  America 135~(1) (2014) 400--406.

\bibitem{gupta2020automatic}
C.~Gupta, E.~Y{\i}lmaz, H.~Li, Automatic lyrics alignment and transcription in
  polyphonic music: Does background music help?, in: ICASSP 2020-2020 IEEE
  International Conference on Acoustics, Speech and Signal Processing (ICASSP),
  IEEE, 2020, pp. 496--500.

\bibitem{sharma2016sonority}
B.~Sharma, S.~M. Prasanna, Sonority measurement using system, source, and
  suprasegmental information, IEEE/ACM Transactions on Audio, Speech, and
  Language Processing 25~(3) (2016) 505--518.

\bibitem{sharma2017vowel}
B.~Sharma, S.~M. Prasanna, Vowel onset point detection using sonority
  information., in: INTERSPEECH, 2017, pp. 444--448.

\bibitem{sharma2018improving}
B.~Sharma, Improving quality of statistical parametric speech synthesis using
  sonority information, Ph.D. thesis (2018).

\bibitem{sharma2014faster}
B.~Sharma, S.~Prasanna, Faster prosody modification using time scaling of
  epochs, in: 2014 Annual IEEE India Conference (INDICON), IEEE, 2014, pp.
  1--5.

\bibitem{rabiner1977use}
L.~Rabiner, On the use of autocorrelation analysis for pitch detection, IEEE
  transactions on acoustics, speech, and signal processing 25~(1) (1977)
  24--33.

\bibitem{tzanetakis2003pitch}
G.~Tzanetakis, A.~Ermolinskyi, P.~Cook, Pitch histograms in audio and symbolic
  music information retrieval, Journal of New Music Research 32~(2) (2003)
  143--152.

\bibitem{Chitra}
C.~Gupta, H.~Li, Y.~Wang, Automatic leaderboard: Evaluation of singing quality
  without a standard reference, IEEE/ACM Transactions on Audio, Speech, and
  Language Processing 28 (2020) 13--26.

\bibitem{grey1978perceptual}
J.~M. Grey, J.~W. Gordon, Perceptual effects of spectral modifications on
  musical timbres, The Journal of the Acoustical Society of America 63~(5)
  (1978) 1493--1500.

\bibitem{sharma2017enhancement}
B.~Sharma, S.~M. Prasanna, Enhancement of spectral tilt in synthesized speech,
  IEEE Signal Processing Letters 24~(4) (2017) 382--386.

\bibitem{vijayan2018analysis}
K.~Vijayan, X.~Gao, H.~Li, Analysis of speech and singing signals for temporal
  alignment, in: IEEE Asia-Pacific Signal and Information Processing
  Association Annual Summit and Conference (APSIPA ASC), 2018, pp. 1893--1898.

\bibitem{sakoe1990dynamic}
H.~Sakoe, S.~Chiba, A.~Waibel, K.~Lee, Dynamic programming algorithm
  optimization for spoken word recognition, Readings in speech recognition 159
  (1990) 224.

\bibitem{cen2011segmentation}
L.~Cen, M.~Dong, P.~Chan, Segmentation of speech signals in template-based
  speech to singing conversion, in: Asia-Pacific Signal and Information
  Processing Association Annual Summit and Conference, 2011, pp. 1--4.

\bibitem{sharma2019combination}
B.~Sharma, H.~Li, A combination of model-based and feature-based strategy for
  speech-to-singing alignment, in: INTERSPEECH, 2019, pp. 624--628.

\bibitem{BLSTM}
L.~Sun, S.~Kang, K.~Li, H.~Meng, Voice conversion using deep bidirectional long
  short-term memory based recurrent neural networks, in: 2015 IEEE
  international conference on acoustics, speech and signal processing (ICASSP),
  IEEE, 2015, pp. 4869--4873.

\bibitem{new2010voice}
T.~L. New, M.~Dong, P.~Chan, X.~Wang, B.~Ma, H.~Li, Voice conversion: From
  spoken vowels to singing vowels, in: IEEE International Conference on
  Multimedia and Expo (ICME), 2010, pp. 1421--1426.

\bibitem{henrich2011vocal}
N.~Henrich, J.~Smith, J.~Wolfe, Vocal tract resonances in singing: Strategies
  used by sopranos, altos, tenors, and baritones, The Journal of the Acoustical
  Society of America 129~(2) (2011) 1024--1035.

\bibitem{oncley1971frequency}
P.~Oncley, Frequency, amplitude, and waveform modulation in the vocal vibrato,
  The Journal of the Acoustical Society of America 49~(1A) (1971) 136--136.

\bibitem{morise2016world}
M.~Morise, F.~Yokomori, K.~Ozawa, {WORLD}: a vocoder-based high-quality speech
  synthesis system for real-time applications, IEICE TRANSACTIONS on
  Information and Systems 99~(7) (2016) 1877--1884.

\bibitem{dehak2010front}
N.~Dehak, P.~J. Kenny, R.~Dehak, P.~Dumouchel, P.~Ouellet, Front-end factor
  analysis for speaker verification, IEEE Transactions on Audio, Speech, and
  Language Processing 19~(4) (2010) 788--798.

\bibitem{wu2016merlin}
Z.~Wu, O.~Watts, S.~King, Merlin: An open source neural network speech
  synthesis system., in: The 9th ISCA Speech Synthesis Workshop (SSW), 2016,
  pp. 202--207.

\bibitem{tokuda2000speech}
K.~Tokuda, T.~Yoshimura, T.~Masuko, T.~Kobayashi, T.~Kitamura, Speech parameter
  generation algorithms for {HMM}-based speech synthesis, in: International
  Conference on Acoustics, Speech, and Signal Processing (ICASSP), Vol.~3,
  2000, pp. 1315--1318.

\end{thebibliography}
}

\end{document}